\documentclass{aa}  
\usepackage{graphicx}
\usepackage{txfonts}
\usepackage{lipsum}
\usepackage{subcaption}         
\usepackage{lscape}             
\usepackage{placeins}           
\usepackage[colorlinks=true, urlcolor=blue, citecolor=blue, linkcolor=blue, breaklinks=true]{hyperref}
\usepackage{xurl}
\usepackage{stfloats}
\usepackage{booktabs}
\usepackage{multirow}
\usepackage{needspace}
\usepackage{makecell}
\usepackage{multicol}
\usepackage{capt-of}
\begin{document}
   \title{Solar energetic electron transport across the heliospheric current sheet}
   \subtitle{A statistical study}
   \titlerunning{Solar energetic electron transport across the heliospheric current sheet}
   \author{C. Han\inst{1,2}\fnmsep\thanks{Corresponding author: chenyaohan@physik.uni-kiel.de}
        \and R. F. Wimmer-Schweingruber\inst{1}
        \and P. Kühl\inst{1}
        \and L. Berger\inst{1}
        \and Z. Ding\inst{1}
        \and A. Kollhoff\inst{1}
        \and Q. Shi\inst{2}
        \and Z. Xu\inst{3}
        \and M. Qin\inst{2}        
        \and M. Wang\inst{4}
        }
   \authorrunning{Han et al.}
   \institute{Institute of Experimental and Applied Physics, Kiel University, 24118 Kiel, Germany
   \and Shandong Key Laboratory of Space Environment and Exploration Technology, School of Space Science and Technology, Institute of Space Sciences, Shandong University, Weihai 264209, China
   \and California Institute of Technology, MC 290-17, Pasadena, CA 91125, USA
   \and Swedish Institute of Space Physics, Box 537, 75121, Uppsala, Sweden
   }
  \abstract
   {Solar eruptive events such as flares and coronal mass ejection (CME)-driven shocks can release solar energetic particles (SEPs) into the heliosphere. The heliospheric current sheet (HCS) is a large-scale structure that separates regions of opposite magnetic polarity. Its influence on SEP propagation remains poorly understood.}
   {In this study, we attempt to provide statistical insight into the role of the HCS in SEP propagation through a statistical analysis of the solar energetic electron (SEE) events observed by Solar Orbiter.}
   {We analysed SEE events in the Comprehensive Solar Energetic Electron event Catalogue (CoSEE-Cat) and classified them into two groups: same-side events, where both the solar source and spacecraft are in the same magnetic sector, and opposite-side events. The magnetic polarities of the solar source region and the spacecraft location are determined comprehensively using approaches based on the potential field source surface (PFSS), magnetic field measurements, the pitch angle distribution of strahl electrons, and the first-order anisotropy of energetic electrons. The uncertainties in determining magnetic polarities associated with these methods are evaluated.}
   {The spacecraft magnetic polarities determined by footpoint positions at the source surface and in situ observations are consistent for most events, providing a useful methodological reference for future studies. We identified 60 same-side events and 9 opposite-side events. By comparing characteristics of events in these two groups, we were able to investigate the potential influence of the HCS on SEP propagation.}
   {Our results show that opposite-side events tend to be more isotropic and that both the solar source and the spacecraft are closer to the HCS than in same-side events. This suggests that particle transport across the HCS is inefficient unless the source or the observer is close to the HCS. These preliminary statistical findings advance our understanding of the role of the HCS in shaping SEP transport.}
   \maketitle
   \nolinenumbers
\section{Introduction}
Solar eruptive events, such as flares and coronal mass ejection (CME)-driven shocks, eject solar energetic particles (SEPs) through the interplanetary magnetic field (IMF) lines into the heliosphere. Understanding SEP propagation is a fundamental topic in heliophysics because it is governed by large-scale magnetic topology, diffusion, drift, and convection, which collectively produce complex spatial and temporal variations in particle intensity and anisotropy \citep[e.g.][and references therein]{van_den_berg_primer_2020}.

The heliospheric current sheet (HCS) is a large-scale structure in the heliosphere \citep{wilcox_quasi-stationary_1965}, formed due to the Sun's dipolar magnetic field, with opposite polarities on either side of the solar equator. The solar wind carries the magnetic field outward, and the HCS appears as a boundary where these opposite polarities meet. The HCS is characterised by a low magnetic field strength, high plasma beta, and reduced solar wind speed compared to the surrounding solar wind \citep{smith_heliospheric_2001}. Due to the tilt of the solar magnetic dipole with respect to the Sun’s rotation axis, the HCS is not flat but twisted into a wavy shape as it is carried outwards by solar wind, also known as the ballerina skirt model \citep{1977RvGSP..15..271A}. 
As a large-scale topological boundary embedded within the solar wind, the HCS represents a natural candidate for affecting SEP transport in the heliosphere. 
Its impact is governed by several distinct physical processes associated with its magnetic structure and plasma environment. The vanishingly weak magnetic field near the neutral sheet causes a breakdown of the adiabatic guiding-centre approximation, leading particles to perform non-adiabatic Speiser orbits composed of meandering motions that facilitate particle injection from the sheet \citep{speiser_particle_1965}. Additionally, the HCS is embedded within a turbulent plasma sheet that promotes strong pitch-angle scattering, which can rapidly isotropise beam-like distributions or act as a diffusive barrier, effectively suppressing cross-hemispheric transport \citep[e.g.][]{winterhalter_heliospheric_1994,smith_heliospheric_2001}. 
Moreover, the polarity reversal across the HCS drives systematic drifts that allow particles to reach longitudes far from their source \citep[e.g.][]{battarbee_modeling_2018,Dalla2020}.

A number of studies report barrier-like or weak HCS effects, suggesting limited cross-sector SEP transport.
\cite{Roelof1973} reported abrupt changes in 0.3 MeV proton fluxes observed by Mariner 5 when the magnetic footpoint of the spacecraft crossed coronal neutral lines 
for the first time, suggesting that crossings of the coronal neutral line by the observer's magnetic footpoint can lead to sudden variations in proton fluxes. 
\cite{Kallenrode1993} conducted the first statistical comparison using Helios 1 and 2 observations of ~0.5 MeV electrons and ~7 MeV protons, showing that event onset and peak times were ordered by the angular distance from the flare to the spacecraft footpoints and were also related to sector boundaries in the source regions.
\cite{shea_comment_1995} reported that most (22 out of 27) ground-level enhancement (GLE) proton events occurred when both the solar active region and the Earth had the same magnetic polarity or either of them was within 10\degr of the HCS, suggesting there might be preferential particle propagation within one polarity structure. 
Statistical studies also reported weak correlations between HCS and SEP. 
\cite{Kahler1996} examined 134 large SEP events and concluded that the streamer structure, the coronal base of the HCS, had no detectable effect on CME-driven shock development. In addition, no significant differences between SEP events originating on the same or opposite side of the HCS relative to the observer. 
\cite{agueda_current_2013} analysed three large near-relativistic electron events (>50 keV) observed in 2001 by both Advanced Composition Explorer (ACE) and Ulysses. 
Their results showed that extended injection profiles occur when spacecraft footpoints lie in the same magnetic sector as the associated flare, whereas sparse and intermittent injections appear when footpoints are in the opposite sector or when warped HCS structures bounded the CME, highlighting the key role of the HCS in shaping SEP injection profiles. 
\cite{battarbee_solar_2017} investigated the effects of a flat HCS on SEP propagation. They found that while significant drifts along the current sheet facilitate transport to distant longitudes, the flat structure itself limits the crossing of particles into the opposite hemisphere. 
\cite{Liou2024} investigated the effect of the HCS on SEP propagation using ~2.0-9.6 MeV/nucleon helium data from Wind. A superposed epoch analysis (SEA) of 319 HCS crossings revealed sharp drops in low-energy (2.00-2.40 MeV/nucleon) SEP fluxes at the HCS, while high-energy (7.40-9.64 MeV/nucleon) fluxes remained largely unaffected. The percentage of flux dropouts decreased with increasing energy, suggesting strong scattering of MeV helium as their gyroradius approaches the HCS thickness. These results provide statistical evidence that the HCS can act as a barrier, suppressing the transport of MeV-energy SEPs from one hemisphere to the other.

In contrast, substantial evidence suggests that the HCS can facilitate SEP transport under specific conditions. 
\cite{battarbee_modeling_2018} extended their simulation work about the flat HCS by incorporating a wavy HCS into their 3D simulations. Their results indicated that the wavy structure allows for more complex particle trajectories, potentially enabling some particles to cross into the opposite hemisphere.
\cite{augusto_relativistic_2019} analysed several GLEs in solar cycle 24 and found that although the source active region was located at the western limb, where it is typically considered magnetically poorly connected to Earth, the presence of the HCS facilitated particle transport. They proposed that when energetic protons are injected within an HCS structure, drift processes enable them to propagate across a wide range of longitudes and reach Earth efficiently. Even for SPEs originating from better-connected regions (e.g. GLEs \#68 and \#69 in January 2005), the HCS does not disrupt the magnetic connectivity; instead, it facilitates it. 
\cite{Waterfall2022} combined statistical analysis of historic GLEs and 3D test particle simulations to highlight the critical role of the HCS in high-energy proton transport. They found that active regions located closer to the HCS (<10\degr) are more likely to be associated with GLEs. Their modelling results showed that proximity to the HCS significantly enhances longitudinal transport and particle intensities at Earth, with the HCS being a relevant factor in reproducing observations for 71\% of the analysed events sampled. 
Even for GeV protons, \cite{moradi_effect_2022} demonstrated that large-scale turbulence enhances the probability of GeV protons reaching the HCS early in an event. This allows particles to utilise the HCS for rapid drift and efficient cross-field transport, even when the injection site is initially separated from the HCS. In a subsequent study regarding different ion species (H$^{+}$, $^3$He$^{2+}$, and $^4$He$^{2+}$), \cite{moradi_impact_2025} found that the resulting intensity profiles are governed by a combination of factors, including the observer's proximity to the HCS, magnetic connectivity, IMF polarity, and drifts dependent on kinetic energy and mass-to-charge ratio ($m/q$). 
Similarly, \cite{DeOliveira2026} analysed the SEP event observed on 1 November 2014, originating from Active Region (AR) 2192 near the western limb. They suggested a plausible scenario involving a combination of solar eruptive activity, particle acceleration by a CME-driven shock, and the presence of an HCS sector crossing, which may have enhanced the magnetic connectivity between the Sun and Earth. 
\cite{Waterfall2025} found that the HCS can facilitate the transport of >10 MeV protons from one hemisphere to the other, based on 3D test particle modelling of three widespread SEP events in 2022–2023, each detected by at least four spacecraft. They employed multiple HCS configurations derived from the Wilcox Solar Observatory, Air Force Data Assimilative Photospheric Flux Transport (ADAPT), and Solar Dynamics  Observatory (SDO)/Helioseismic and Magnetic Imager (HMI) (via Predictive Science Inc., PSI) and compared the modelled proton flux profiles with in situ measurements at each observer location. Their results indicate that inclusion of the HCS is essential to reproduce both the wide longitudinal spread of particles and the observed flux profiles, with the highest fluxes occurring at observers located nearest to the HCS. These findings demonstrate the critical role of the HCS in shaping the large-scale transport and longitudinal distribution of energetic protons, consistent with prior drift-based studies \citep[e.g.][]{levy_interplanetary_1976,burger_drift_1985,Dalla2020}.

Taken together, the previous studies present a fundamentally mixed picture in which the HCS has been identified as a magnetic barrier, a conduit, or a neutral structure in SEP transport, depending on the particle species, particle energy, and the distribution of the source region. The lack of consensus indicates that the role of the HCS in SEP transport remains unresolved and that further investigation is warranted. This study investigates the influence of the HCS on the propagation of SEEs and uses the Comprehensive Solar Energetic Electron event Catalogue \citep[CoSEE-Cat;][]{Warmuth2025}. We classified SEE events into two groups: same-side events, where both the solar source and spacecraft are in the same magnetic sector, and opposite-side events, where they are in opposite sectors. Section \ref{sec:event_catalogue} introduces the SEE catalogue used in this study. Section \ref{sec:Event_classification_method} details the classification methods for events based on their location relative to the HCS. The classification results are presented in Section \ref{Results}, followed by a discussion in Section \ref{Discussion}.
\section{Event catalogue and instrumentation}
\label{sec:event_catalogue}
The SEE events used in this study are sourced from the Comprehensive Solar Energetic Electron event Catalogue (CoSEE-Cat) \citep{Warmuth2025}. The CoSEE-Cat catalogue is a comprehensive catalogue of SEE events identified by the Energetic Particle Detector \citep[EPD;][]{Rodriguez-Pacheco2020} suite onboard Solar Orbiter. It provides basic parameters for all observed SEE events, as well as contextual information on associated solar phenomena, including X-ray flares detected by the Spectrometer/Telescope for Imaging X-rays \citep[STIX;][]{Krucker2020a}, EUV eruptions by the Extreme Ultraviolet Imager \citep[EUI;][]{Rochus2020a}, radio bursts by the Radio and Plasma Waves instrument \citep[RPW;][]{Maksimovic2020a}, and CMEs by Metis \citep{Antonucci2020a} and the Solar Orbiter Heliospheric Imager \citep[SoloHI;][]{Howard2020a}. Additionally, it includes data on interplanetary conditions near the spacecraft measured by the Solar Wind Analyser (SWA)/Proton and Alpha particle Sensor (PAS) \citep{Owen2020} and the Magnetometer \citep[MAG;][]{Horbury2020}. At the time of writing, the CoSEE-Cat catalogue contains 303 SEE events observed from November 2020 through December 2022.

Apart from the catalogue, in this study, we also used in situ measurements from Solar Orbiter. These include electron measurements from the Electron Proton Telescope (EPT), which is part of EPD; solar wind speed measurements from the SWA; solar wind electron measurements from the SWA/Electron Analyser System (EAS); and magnetic field measurements from the MAG.
\section{Event classification method}
\label{sec:Event_classification_method}
To investigate the function of the HCS during the propagation of SEPs, we divided the events listed in the CoSEE-Cat catalogue into two groups: same-side events and opposite-side events. Specifically, the same-side events are defined as those events where the solar source and the spacecraft are located in the same magnetic sector, while opposite-side events are those for which the solar source and spacecraft are located in opposite sectors. To classify the events accordingly, we needed to determine which magnetic sector contains the associated solar source and the spacecraft during each event. The procedures for obtaining this information are described in the following sections.
\subsection{Identification of the solar source magnetic sector}
\label{sec:Identification_of_the_solar_source_magnetic_sector}
In the CoSEE-Cat catalogue, \cite{Warmuth2025} identified the solar sources of SEE events by associating them with multiple solar phenomena, including solar flares, type III radio bursts, and CMEs. Their analysis reveals that over 88\% of SEE events in the CoSEE-Cat catalogue are associated with X-ray flares observed by STIX, making these flares the primary candidates for the solar sources of the SEE events. Therefore, in this study, we considered the X-ray flares observed by STIX as the solar sources of the SEE events. The STIX flares' locations at the photosphere were reconstructed using the Expectation Maximization imaging algorithm \citep[EM;][]{massa_count-based_2019} and provided in the CoSEE-Cat catalogue.

To get the flare location relative to the HCS, we carried out a potential field source surface extrapolation (PFSS) to obtain the HCS position at the source surface (SS) for each event. The PFSS extrapolation is a widely utilised technique for modelling the magnetic field geometry in the solar atmosphere \citep{Schatten1969}. This method uses a photospheric magnetogram as the inner boundary condition for the magnetic field potential, which must satisfy the Laplace equation. The model assumes that the magnetic field becomes purely radial at an outer boundary, termed the SS. The SS is a theoretical spherical boundary. Inside the SS, the solar magnetic field is complex, with both closed and open field lines, whereas beyond the SS, the magnetic field lines are assumed to be basically open, stretching radially outwards with the solar wind. Ideally, the HCS forms a stable structure only outside the SS. The SS is typically set at $2.5$ R$_\odot$ (solar radii), originally suggested by \cite{Altschuler1969b}, and we also adopted this value for the present study. For the implementation, we utilised the open-source \texttt{sunkit-magex}\footnote{\url{https://github.com/sunpy/sunkit-magex}} package, which is fully integrated in \texttt{sunpy} \citep{Stansby2020}.

Magnetograms serve as boundary conditions for coronal magnetic field models and, thus, they are essential inputs for the PFSS. Prior research demonstrates that the selection of the magnetogram source can substantially influence the resulting coronal magnetic field configuration and, consequently, the derived HCS structure \citep[e.g.][]{Perri2023,Koukras2025}. In this work, we utilised publicly available magnetograms from three sources: HMI, Global Oscillation Network Group (GONG), and ADAPT. To evaluate the sensitivity of our results to the choice of magnetogram, we compared the HCS locations obtained from each source. Each source offers multiple versions of magnetograms, reflecting different data-processing and assimilation methods. For consistency and reliability, we selected the most recent and widely adopted versions. Specifically, for HMI, we used the polar-corrected daily synoptic map \texttt{hmi.mrdailysynframe\_polfil\_720s} (available at Joint Science Operations Center (JSOC)\footnote{\url{http://jsoc.stanford.edu/}}), which is typically used in recent research \citep[e.g.][]{Milanovic2023,Heinemann2024,Heinemann2025}. The HMI polar-corrected daily synoptic map used in this study is not a full Carrington-rotation (CR) synoptic map; instead, it is in a synchronic frame and better represents the front-side conditions. Data within the central meridian (CM) $\pm 60\degr$ longitudinal window in the synoptic map are replaced by the daily full-disk observation at the corresponding date \citep{Sun2018,Perri2023}, while the remaining regions are filled with standard synoptic-map data. The more up-to-date version of the synoptic map is expected to reduce the time gap between the data and the event time and to better capture fast-evolving structures. Following a similar idea, for GONG, we chose the GONG-zqs (zeropoint-corrected) hourly synoptic maps (available at the National Solar Observatory's data archive\footnote{\url{https://gong.nso.edu/data/magmap/QR/zqs/}}), which are also widely used \citep[e.g.][]{Wang2025,Kumar2025} and officially recommended\footnote{\url{https://gong2.nso.edu/archive/patch.pl?menutype=zeroPoint\#step2}}. The ADAPT magnetogram combines photospheric magnetic field observations from the GONG telescope network with the ADAPT flux transport model \citep{worden_evolving_2000}. This data assimilation allows for the construction of full synchronic maps, providing a complete global view of the magnetic field at the time of observation. The ADAPT magnetogram provides twelve realisations at the time of observation, based on different assimilation techniques \citep{Hickmann2015}. To address the challenge of selecting the optimal realisation, we employed the Magnetic Connectivity Tool\footnote{\url{http://connect-tool.irap.omp.eu/}} \citep{Rouillard2020}, which automatically identifies the best match for each event by comparing coronal and heliospheric models with white-light (WL) observations \citep{Poirier2021}. For all three magnetograms, we chose the magnetogram whose time is closest to the solar release time (SRT) determined by the time shift analysis (TSA) method \citep{Vainio2013,paassilta_catalogue_2018} for each event. The SRT calculated by TSA is provided in the CoSEE-Cat catalogue and is available for all events. The HCS at the SS is identified as the magnetic polarity inversion line (PIL) in the radial magnetic field component ($B_r$) maps.
\begin{figure}[!t]
\centering
\includegraphics[width=\hsize]{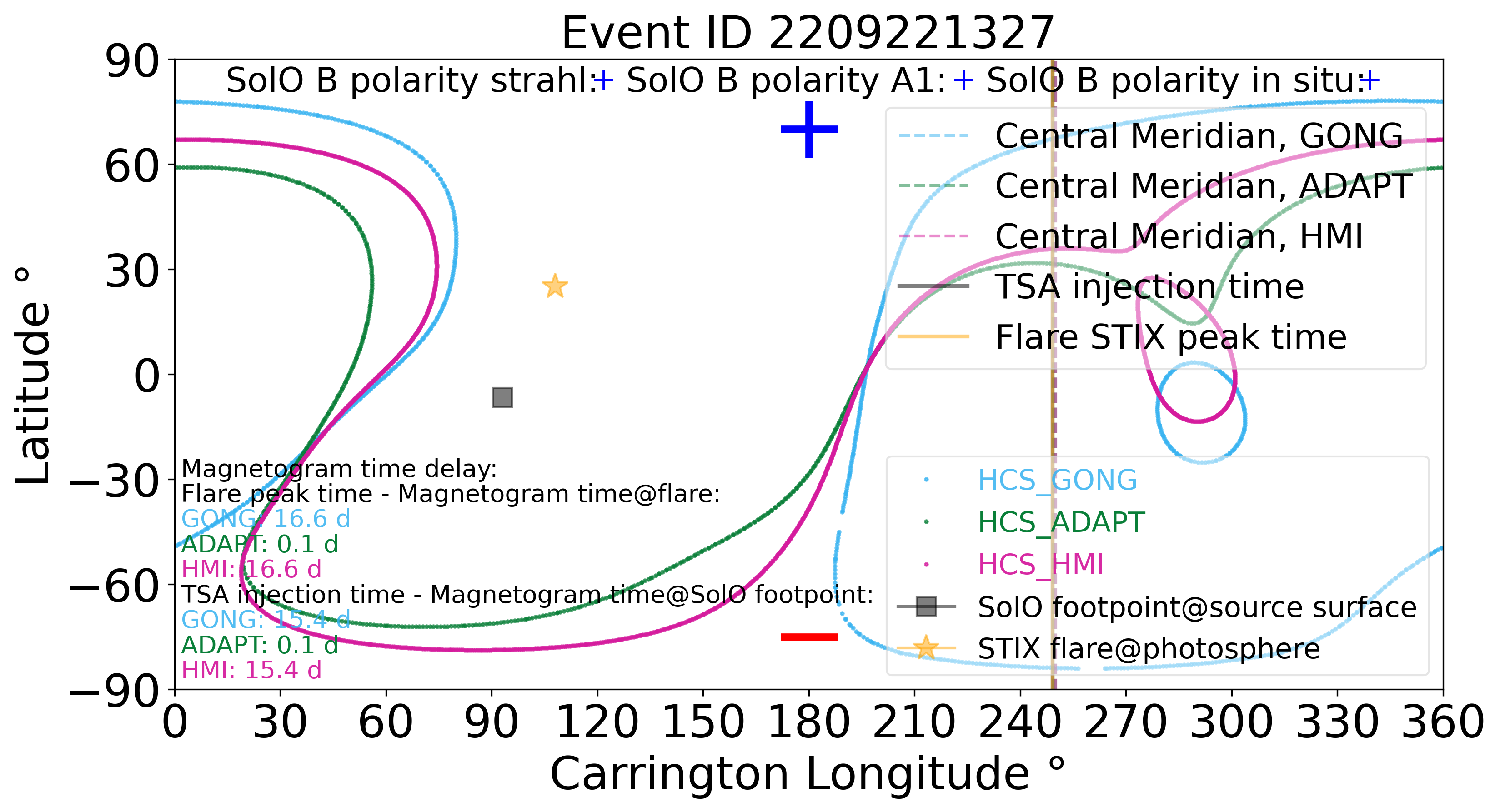}
\caption{Example HCS position plot of the event \#2209221327. The unique event identifier encodes the EPD onset time, corresponding here to 2022 Sep 22 at 13:27 UT. The cyan, green, and magenta lines represent the HCS positions derived from GONG, ADAPT, and HMI magnetograms, respectively. The yellow star indicates the flare at the photosphere, and the black square denotes the Solar Orbiter's footpoint at the SS. The magnetic polarities at Solar Orbiter determined by other methods are shown at the top of each panel, with outwards indicated by blue plus signs and inwards by red minus signs. The dashed lines in the corresponding colours mark the CM longitude positions for each magnetogram. The black and yellow solid lines represent the longitude of the event's injection time determined by the TSA method and flare peak time in the synoptic map, respectively, both provided in the CoSEE-Cat catalogue. The time latencies of each magnetogram relative to the flare and Solar Orbiter's footpoint are also indicated in the bottom left corner of the figure.}\vspace{-4mm}
\label{HCS_example}
\end{figure}

In this study, the solar source magnetic sector was identified using the radially projected flare position. This choice was motivated by the fact that PFSS mapping of flare locations from photosphere to the SS is often non-unique and sensitive to the adopted magnetogram and PFSS setup, while the PFSS-derived flare footpoints generally have the same magnetic polarity as the radially projected flare position. 
A more detailed discussion is provided in Appendix~\ref{appendix_Methodological_uncertainties_and_limitations}. 
An example of the HCS derived from different magnetogram sources is shown in Fig.~\ref{HCS_example}. The HCS locations derived from GONG, ADAPT, and HMI magnetograms are denoted by cyan, green, and magenta lines, respectively. The yellow star indicates the flare location at the photosphere associated with event \#2209221327, which indicates an 'outwards' magnetic polarity for all three magnetograms. 
To enhance the reliability of determining the flare's magnetic sector, we only selected events where the flare's magnetic sector is consistent across all three magnetograms.
\subsection{Identification of the spacecraft magnetic sector}
\label{sec:Identification_of_the_spacecraft_magnetic_sector}
The dynamic nature of the solar wind and magnetic field fluctuations can complicate the accurate determination of the spacecraft magnetic sector. 
Therefore, we employed multiple methods to enhance the reliability of our determination; only events for which the spacecraft magnetic sector is consistently classified across all four methods listed below (i.e. in situ magnetic field measurements, ballistic back-mapping, pitch angle distribution (PAD) of strahl electrons, and the
first-order anisotropy of energetic electrons) are included in the further analysis.
\subsubsection{In situ magnetic field measurements}
Using the in situ magnetic field measurements from MAG onboard Solar Orbiter might be the most straightforward way to determine the spacecraft magnetic sector. We calculated the angle between the local magnetic field vector and the nominal Parker spiral direction away from the Sun, as shown in the fourth panel of Fig.~\ref{time_profile_example}. If this angle is less than 90\degr, the magnetic field polarity is classified as 'outwards', and if it is greater than 90\degr, it is classified as 'inwards'. Specifically, we calculated the angle within 20 minutes around the onset time and used the predominant polarity to determine the magnetic polarity of each event. For events without solar wind speed measurement, we considered a range of solar wind speeds from 200 km/s to 1000 km/s to calculate the angle and choose the solar wind speed with the median angle closest to 90\degr. The example shown in the fifth panel of Fig.~\ref{time_profile_example} exhibits angles all less than 90\degr, indicating an 'outwards' magnetic polarity at the spacecraft location.
\subsubsection{Ballistic back-mapping}
With the position of the HCS in hand, a classic method employed to determine the spacecraft magnetic sector is ballistic back-mapping from the spacecraft to the SS, assuming a nominal Parker spiral and constant solar wind speed, $v_{\rm sw}$, expressed as
\begin{equation}
   \varphi(r) = \varphi_0 + \frac{\omega(\vartheta)}{v_{\mathrm{sw}}} \cdot (r_0 - r_{SS}) \cdot \cos\vartheta
\end{equation}
where $\varphi$ is the longitude of Solar Orbiter's footpoint at the SS, $\varphi_0$ is the longitude of Solar Orbiter, $v_{\mathrm{sw}}$ is the solar wind speed measured in situ by SWA onboard Solar Orbiter, $r_0$ and $r_{SS}$ are the radial distances of Solar Orbiter and the SS, respectively, and $\omega$ is the differential solar rotation frequency at different heliographic latitude $\vartheta$ \citep{Gieseler2022}. The latitude of Solar Orbiter's footpoint is equal to Solar Orbiter's latitude. When the solar wind speed measurement was not available, we considered a set of solar wind speeds from 200 km/s to 1000 km/s to perform the back-mapping and only took those events for which all the footpoints were on the same side of the HCS into further analysis. This procedure was not intended to reject any particular class of solar wind conditions, but rather to exclude events whose magnetic sector identification is highly sensitive to the mapping assumptions. In such cases, the event remains ambiguous without a reliable in situ solar wind speed measurement. As an example, Figure~\ref{HCS_example} shows the Solar Orbiter footpoint location at the SS for event \#2209221327, derived from ballistic back-mapping, and marks the footpoint with a black square. In this event, the spacecraft is located in an outwards magnetic polarity sector.
\subsubsection{Pitch angle distribution of strahl electrons}
\begin{figure}[!t]
\centering
\includegraphics[width=\hsize]{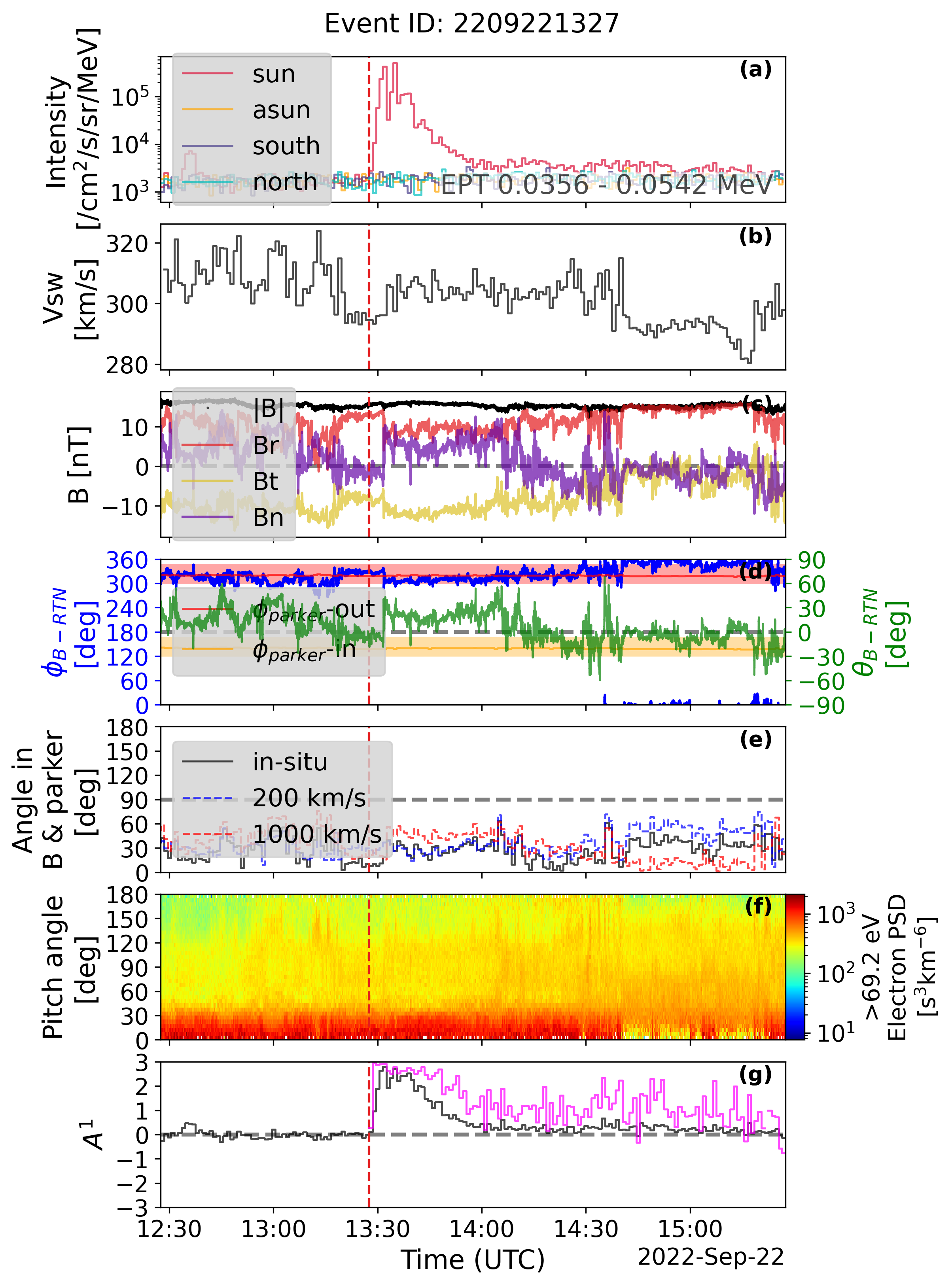}
   \caption{Example plot of the event \#2209221327. The vertical red dashed line denotes the event onset time. From top to bottom the panels display: (a) electron intensities in the 35.6-54.2 keV measured by the four EPT telescopes; (b) solar wind speed observed by SWA on board Solar Orbiter; (c) magnetic field strength, with component colours radial (red), tangential (yellow), normal (purple) and total (black); (d) magnetic field azimuthal (blue) and latitudinal (green) angles, along with the azimuthal angles of outwardly vs inwardly (red vs orange) directed Parker spiral field lines calculated using solar wind speed from in situ observations (solid lines) and assumptions from 200 km/s to 1000 km/s (shadow regions); (e) the angle between the magnetic field vector and the nominal Parker spiral direction away from the Sun; (f) the PAD of strahl electrons measured by SWA/EAS; and (g) the first-order anisotropy $A^1$ for the measured electron intensities, shown both without (black) and with (magenta) background subtraction.}
      \label{time_profile_example}
\end{figure}

Strahl electrons are a population of suprathermal electrons that originate in the solar corona \citep{Feldman1975a,Rosenbauer1977a} and travel along magnetic field lines, appearing at energies above ~50 eV \citep{Gosling1987a}. As the strahl electrons typically move away from the Sun, the PAD of the strahl electrons is usually used to determine the magnetic polarity at the spacecraft location \citep[e.g.][]{owens_solar_2013,Owens2017,Owen2022}. In this study, electrons with energies $\geq$ 70 eV \citep{Owen2022} measured by the SWA/EAS onboard Solar Orbiter were used to represent the strahl electrons. As an example, the sixth panel of Fig.~\ref{time_profile_example} shows the PAD of strahl electrons during event \#2209221327, which exhibits a clear concentration near $0^\circ$. 
To determine the strahl electron direction, we employed the simple algorithm of \citep{Owens2017}, which classifies the magnetic polarity based on the relative magnitudes of the parallel and antiparallel electron phase space densities (PSD) compared with the background PSD. The parallel electron population is defined as electrons with pitch angles between 0\degr and 45\degr, while the antiparallel population is between 130\degr and 180\degr. Electrons with pitch angles between 65\degr and 115\degr are used to represent the background. If the parallel or antiparallel PSD exceeds the background level by some threshold, a parallel or antiparallel strahl is determined to exist. In this study, we required the median of the averaged parallel or antiparallel electron PSD within 20 minutes around the onset time of the SEE event to be 30\% above the background to identify strahl electrons. We further required counterstreaming electrons to have parallel and antiparallel PSD within 30\% of each other to avoid overclassification of counterstreaming electrons. As strahl electrons generally move away from the Sun, the presence of parallel strahl electrons implies an 'outwards' magnetic polarity, whereas antiparallel strahl electrons indicate an 'inwardly' directed polarity. The example shown in Fig.~\ref{time_profile_example} for event \#2209221327 exhibits clear parallel strahl electrons, indicating an 'outwards' magnetic polarity at the spacecraft location.
\subsubsection{The first-order anisotropy}
\label{sec:first-order_anisotropy}
The first-order anisotropy, $A^1$, quantifies the degree of anisotropy in the particle distribution, with positive values indicating a predominance of particles moving parallel to the magnetic field and negative values indicating antiparallel. Similarly to the strahl electrons, we assume that the electrons observed by EPT propagate away from the Sun. Therefore, if the first-order anisotropy $A^1$ is positive, the magnetic polarity is classified as 'outwards', and if it is negative, it is classified as 'inwards'. Specifically, in this study, the predominant sign of the first-order anisotropy $A^1$ in the 10 minutes after the onset time was used to determine the magnetic polarity of each event. 
We used the four sector intensities observed by EPT and the measurements of the local magnetic field vectors measured by MAG onboard Solar Orbiter to calculate the first-order anisotropy $A^1$ for each event using the following equation proposed by \cite{Brudern2018} and expressed as
\begin{equation}
A^1=3\frac{\sum_{\mathrm{i}=1}^4\delta\mu_i\cdot\mu_i\cdot I\left(\mu_i\right)}{\sum_{\mathrm{i}=1}^4\delta\mu_i\cdot I\left(\mu_i\right)},
\end{equation}
where $\mu_i$ is the central pitch angle cosine of the $i$th telescope, $\delta\mu_i$ is the pitch angle cosine range of the telescope opening cone, and $I(\mu_i)$ is the intensity measured in the $i$th telescope.

As concluded in \citep{Warmuth2025}, SEE event anisotropies in the CoSEE-Cat catalogue are often underestimated without a proper background subtraction. We also performed a background subtraction before calculating the first-order anisotropy $A^1$, where the background is defined simply as the average intensity within 1 hour before onset time. We did not employ the complex background subtraction method detailed in \citep{Warmuth2025}, as our primary goal in this study is to determine the sign of $A^1$ to infer magnetic polarity, rather than to perform a precise quantitative comparison of anisotropy values. 
The bottom panel in Fig.~\ref{time_profile_example} shows an example of the first-order anisotropy $A^1$ for the measured electron intensities, plotted both without (black) and with (magenta) background subtraction, which exhibits a positive $A^1$, indicating an 'outwardly' directed magnetic polarity at the spacecraft location.
\section{Results}
\label{Results}
\subsection{Identification results for solar source and spacecraft magnetic sectors}
\begin{figure*}[tbp]
\centering
\includegraphics[width=\hsize]{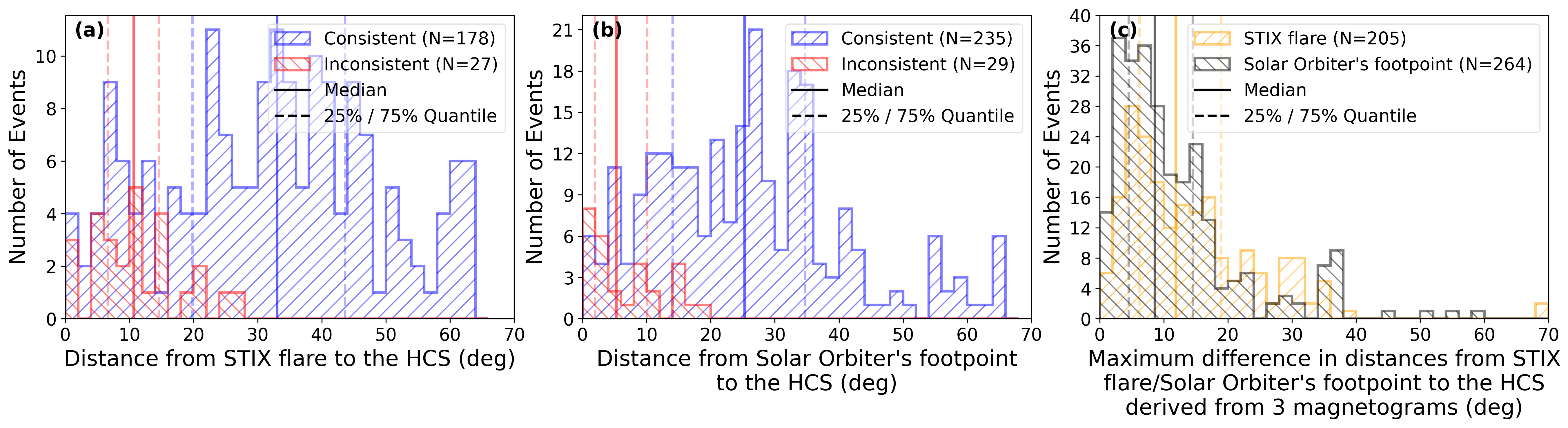}
   \caption{Distribution of the minimum distance from flare (a) and Solar Orbiter's footpoint (b) to the HCS at the SS. The events where the flare/Solar Orbiter's footpoint location relative to the HCS is consistent across all three magnetograms are shown in blue, ambiguous ones in red. The minimum distances from flare and Solar Orbiter's footpoint to the HCS in (a) and (b) are calculated based on the HCS derived from ADAPT magnetogram. (c) The distribution of the maximum difference in the distance from the flare (yellow bars) and  Solar Orbiter's footpoint (black bars) to the HCS as derived from three magnetograms. Solid lines in corresponding colours represent the median value and the dashed lines represent the 25\%/75\% quantile.}
      \label{distance_HCS_3MGs}
\end{figure*}

As mentioned in \ref{sec:Identification_of_the_solar_source_magnetic_sector}, in this study, we projected the flare location from the photosphere radially to the SS. Based on the HCS position derived at the SS, we calculated the minimum angular distance between the flare and the HCS. This calculation assumes a spherical SS, where the distance is defined as the central angle between the projected flare coordinates and the nearest point on the HCS. The distribution of the minimum distance from the flare to the HCS for all events is shown in Fig.~\ref{distance_HCS_3MGs} (a). The blue bars represent events for which the flare location relative to the HCS is consistent across all three magnetograms, and the red bars show those with inconsistent results. It is clear that events for which the flare location relative to the HCS is inconsistent across the three magnetograms tend to be closer to the HCS.
Similarly to the flare situation, we calculated the minimum distance from Solar Orbiter's footpoint to the HCS for all events, as shown in Fig.~\ref{distance_HCS_3MGs} (b). Events with inconsistent classifications across the magnetograms typically involve footpoints located closer to the HCS, mirroring the trend observed for flare locations.

Furthermore, for each event, we quantified the discrepancy in the HCS location by calculating the maximum difference in the distances from the flare and Solar Orbiter's footpoint to the HCS, as derived from three magnetograms. The distribution of these maximum differences is presented in Fig.~\ref{distance_HCS_3MGs} (c), where the red bars correspond to the flare and the black bars to Solar Orbiter's footpoint. It is evident that for the majority of events, the maximum difference remains below 20\degr. This value is consistent with the upper limit of the flare-to-HCS and Solar Orbiter's footpoint-to-HCS distances observed for events with inconsistent magnetic sector determinations across the three magnetograms (see Fig.~\ref{distance_HCS_3MGs} (a) and (b)). This finding suggests that the inconsistencies in determining the magnetic sector for a flare or Solar Orbiter's footpoint located near the HCS are likely primarily attributable to uncertainties in reconstructing the HCS position using different magnetograms.

\begin{figure}
\centering
\includegraphics[width=\hsize]{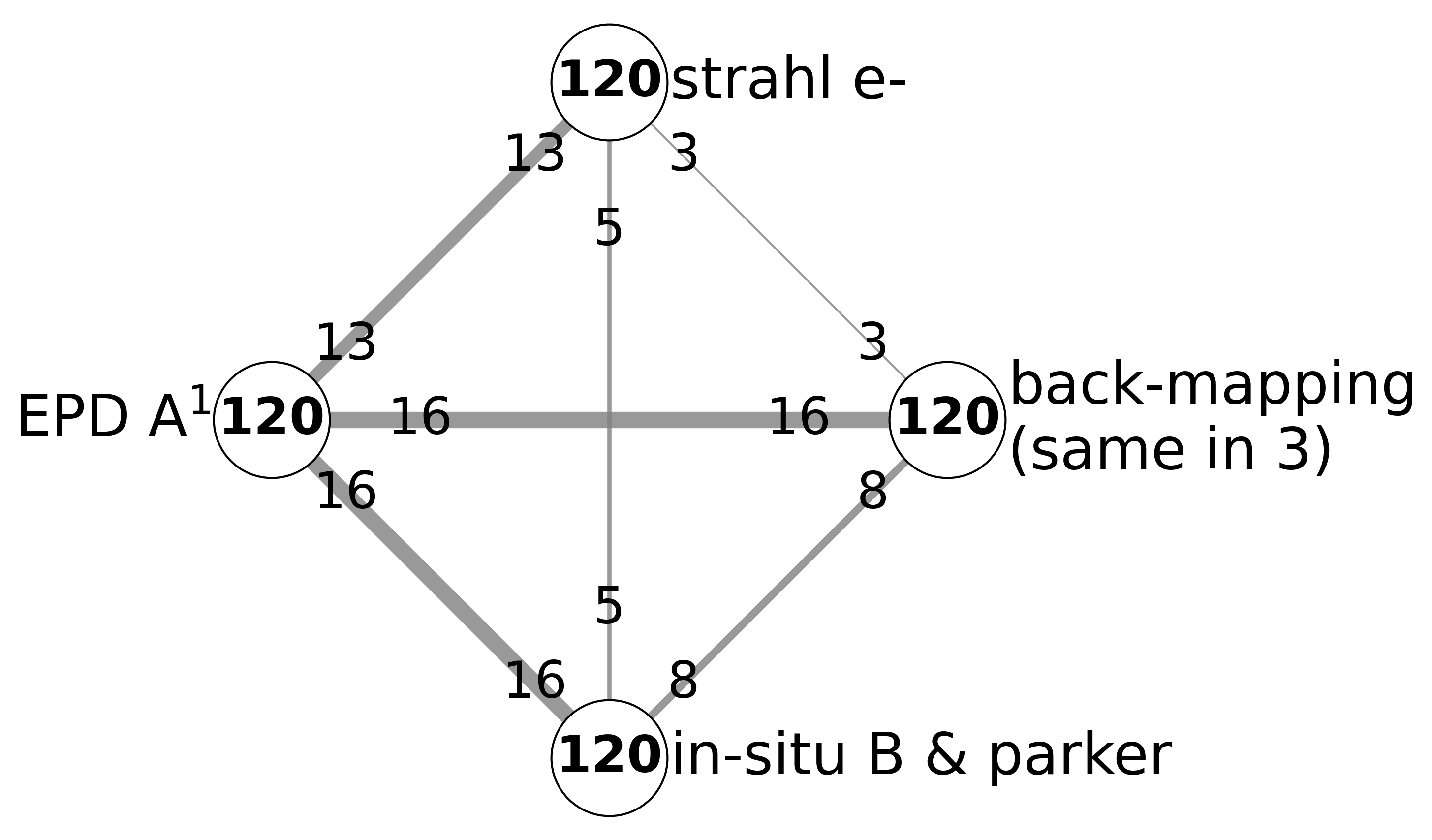}\vspace{-3mm}
   \caption{Comparison of the Solar Orbiter magnetic sectors determined by four methods: ballistic back-mapping, methods based on PAD of strahl electrons, in situ magnetic field measurements, and the first-order anisotropy $A^1$ of EPT electrons. Numbers in circles indicate the count of events with available data for all methods. The thickness of the lines and numbers on the lines represent the number of events with inconsistent polarity determinations between two methods.}\vspace{-5mm}
      \label{SolO_B_polarity_comp4}
\end{figure}

In Fig.~\ref{SolO_B_polarity_comp4}, we compare the Solar Orbiter magnetic sector determined by four methods: ballistic back-mapping, methods based on PAD of strahl electrons, in situ magnetic field measurements, and the first-order anisotropy $A^1$ of EPT electrons. For ballistic back-mapping methods, we only kept those events where the magnetic sector determination is consistent across all three magnetograms. The numbers in circles indicate the count of events with available data for all methods, for a total of 120 events. The thickness of the lines and numbers on the lines represent the number of events with inconsistent magnetic sector determinations between the two methods. It is evident that most events exhibit consistent magnetic sector determinations across all four methods. The comparison reveals that the method using PAD of strahl electrons shows the highest consistency with the other three methods. Notably, only 3 out of 120 events show inconsistencies between the method based on PAD of strahl electrons and the ballistic back-mapping method. The method based on the first-order anisotropy $A^1$ of EPT electrons shows the most discrepancies when compared to the other methods. This likely reflects both the complex transport history of energetic electrons and the observational limitations of the anisotropy measurement itself. During the transit from the Sun to the spacecraft, energetic electrons undergo pitch-angle scattering due to interactions with magnetic field fluctuations and turbulence. Strong scattering can rapidly isotropize the electron distribution, reducing the $|A^1|$ and making the determination ambiguous \citep[e.g.][]{wibberenz_multi-spacecraft_2006, zhang_propagation_2009}. Perpendicular transport or diffusion can further weaken the directional signature of the electrons, so that a small $|A^1|$ does not necessarily reflect the local magnetic polarity alone. Additionally, the presence of large-scale magnetic structures in the heliosphere, such as a stream interaction region (SIR) or interplanetary coronal mass ejection (ICME) located beyond the spacecraft can reflect outwardly propagating electrons back toward the Sun, creating bidirectional flows \citep[e.g.][]{tan_observational_2009}. Moreover, the interpretation of $A^1$ is limited by the pitch-angle coverage of EPT, because restricted sampling of pitch-angle space can also produce an ambiguous anisotropy even when the electrons are not truly isotropic. In such cases, the observed anisotropy may not accurately reflect the local magnetic field polarity, leading to the discrepancies observed in our statistical comparison.
\subsection{Event classification results}
\label{sec:Event_classification_results}
As mentioned above, to enhance the reliability of determining the flare magnetic sector, we only included those events for which the flare magnetic sectors are consistent across all three magnetograms. For the spacecraft, we employed four methods to determine its magnetic sector and only retained those events where the classification is consistent across all methods. We only considered events with available data for all methods listed above, resulting in a reduced final sample. Finally, 
we identified a total of 69 events, comprising 60 same-side events and 9 opposite-side events. 
A list of events with same- and opposite-side identifications is given in Appendix ~\ref{appendix_event_list}. 
Same-side events significantly outnumber opposite-side events.

\begin{figure*}[tbp]
\centering
\includegraphics[width=\textwidth]{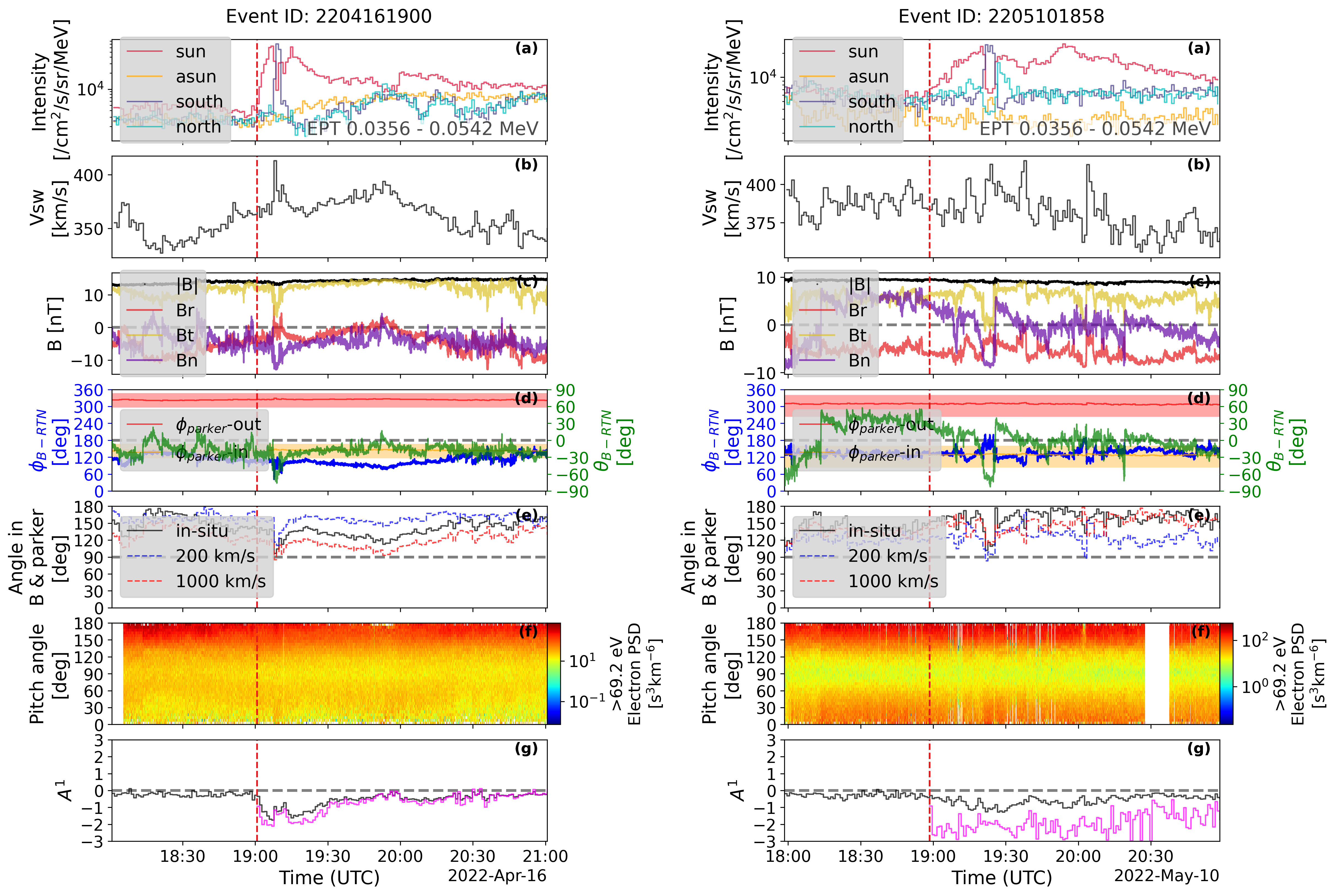}
   \caption{Time profiles of two example events: a same-side event (event \#2204161900, left panel) and an opposite-side event (event \#2205101858, right panel). The indications of each panel are the same as those in Fig.~\ref{time_profile_example}.}
      \label{time_profile_examples}
\end{figure*}

\begin{figure*}[tbp]
\centering
\includegraphics[width=\textwidth]{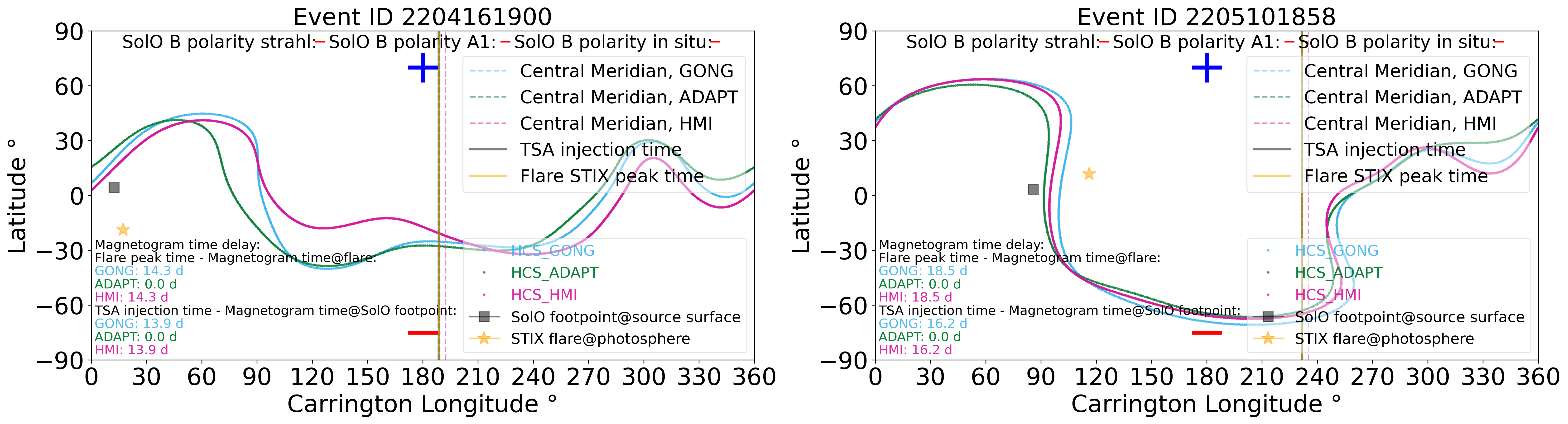}
   \caption{HCS position at the SS during two example events: a same-side event (event \#2204161900, left panel) and an opposite-side event (event \#2205101858, right panel). The indications of each element are the same as those in Fig.~\ref{HCS_example}.}
      \label{HCS_examples}
\end{figure*}

As illustrative examples, we present a same-side event \#2204161900 and an opposite-side event \#2205101858. Figure~\ref{time_profile_examples} displays the time profiles for both events, with the panel format identical to that in Fig.~\ref{time_profile_example}. Figure~\ref{HCS_examples} illustrates the positions of the HCS, the flare, and the Solar Orbiter's footpoints during these events, along with the Solar Orbiter's magnetic polarity results determined by the alternative methods displayed at the top of the figure. 
Additionally, the longitude of the CM for each magnetogram is indicated by a dashed line in the corresponding colour. The black and yellow solid vertical lines represent the longitude of the particle injection time determined by the TSA method and the flare peak time in the synoptic map, respectively, both provided in the CoSEE-Cat catalogue. The time latencies of each magnetogram relative to the flare and Solar Orbiter's footpoint are provided in the bottom-left corner. For GONG and HMI magnetograms, as detailed in ~\ref{sec:Identification_of_the_solar_source_magnetic_sector}, we employed synchronic-frame synoptic maps. These are composite maps that utilise daily full-disk observations within $\pm 60^{\circ}$ of the CM and standard synoptic-map data for the remaining regions. Consequently, if a flare or footpoint is located within $\pm 60^{\circ}$ of the CM, the magnetogram time is taken as the CM time; otherwise, it corresponds to the time of that longitude in the full CR synoptic map. By contrast, as ADAPT magnetograms are fully synchronic maps, the time for all locations is identical to the CM time. The latency of the magnetogram relative to the flare is calculated as the difference between the flare peak time and the magnetogram time at the flare's location; similarly, the latency relative to Solar Orbiter's footpoint is the difference between the particle injection time and the magnetogram time at the footpoint's location.

\begin{figure*}[tbp]
\centering
\includegraphics[width=\textwidth]{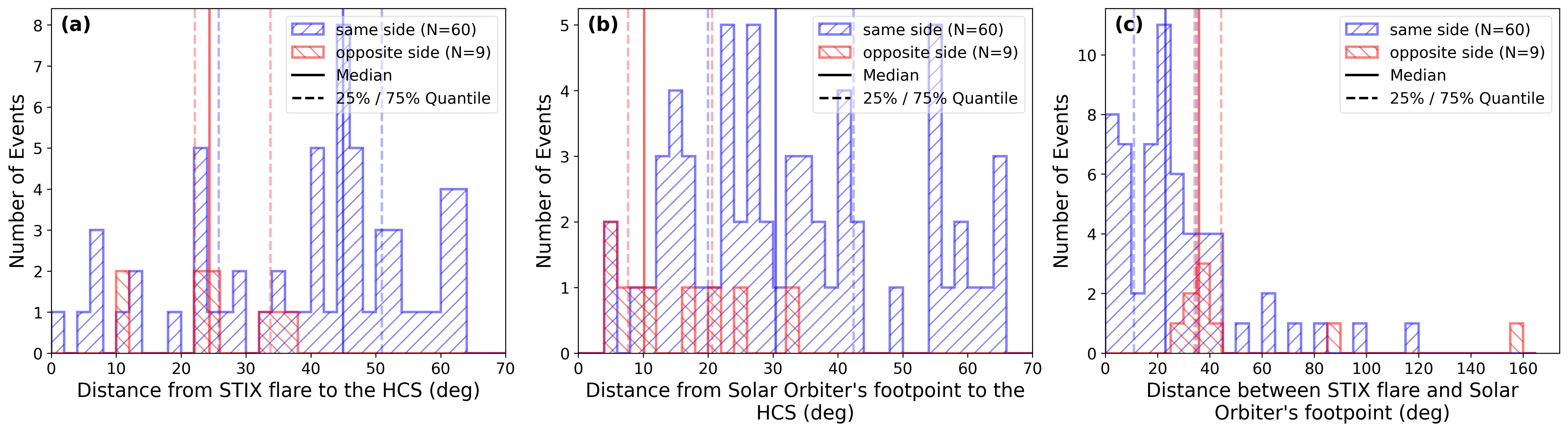}
   \caption{Distributions of distances from flare to the HCS (a), from Solar Orbiter's footpoint to the HCS (b) and from flare to Solar Orbiter's footpoint (c) of the same-side events (blue bars) and opposite-side events (red bars).}
      \label{angle_diff_HCS_side}
\end{figure*}

We calculated the minimum distance from both the flare and Solar Orbiter's footpoint to the HCS for same-side events and opposite-side events, as shown in Fig.~\ref{angle_diff_HCS_side}. The left panel displays the distribution of the flare-to-HCS distance, while the middle panel shows the distribution of Solar Orbiter's footpoint-to-HCS distance. In both panels, the blue bars represent same-side events, and the red bars denote opposite-side events. It is evident that both the flare and Solar Orbiter's footpoint in opposite-side events tend to be closer to the HCS compared to those in same-side events. This finding suggests that proximity to the HCS may facilitate magnetic connectivity between the solar source and the spacecraft during SEP events. Specifically, for opposite-side events, it appears that sufficient magnetic connectivity is established only when the flare or the spacecraft is in close proximity to the HCS. Conversely, if both of them are located far from the HCS, the likelihood of observing an SEP event diminishes significantly due to the lack of efficient transport pathways across the sector boundary. As demonstrated in \citep{Waterfall2025}, SEP access is strongly dependent on an observer's magnetic proximity to the HCS, not just to the eruption site. Our statistical results further substantiate this perspective, suggesting that proximity to the HCS is a prerequisite for the observability of opposite-side events, likely because the HCS region serves as a critical corridor for cross-sector particle transport. 
Interestingly, the finding that both flares and Solar Orbiter's footpoints in opposite-side events tend to be closer to the HCS aligns with the results of \cite{shea_comment_1995}. They found that 8 out of 27 GLEs have apparently different magnetic field polarity domains, and among them, 5 events have the source region or the Earth's footpoint located within 10\degr\ of the HCS.

The right panel of Fig.~\ref{angle_diff_HCS_side} illustrates the distribution of the distance between the flare and Solar Orbiter's footpoint for same-side events (blue bars) and opposite-side events (red bars). It seems that no opposite-side event has a distance between the flare and Solar Orbiter's footpoint smaller than 20\degr. This result may not be physical due to the uncertainty in reconstructing the HCS position using different magnetograms. As shown in Fig.~\ref{distance_HCS_3MGs} (c), the maximum difference in the distance from the flare (or Solar Orbiter's footpoint) to the HCS as derived from the three magnetograms mostly remains below 20\degr. Consequently, if a flare or Solar Orbiter's footpoint lies in close proximity to the HCS, it is highly probable that its position relative to the HCS will be inconsistent across different magnetograms due to variations in the modelled HCS location. As a result, such events will be excluded from our analysis, as described in ~\ref{sec:Identification_of_the_solar_source_magnetic_sector}. This selection effect may therefore explain why no opposite-side event is found with a flare--footpoint separation smaller than 20\degr.

Also as shown in Fig.~\ref{angle_diff_HCS_side} (c), the majority of the events exhibit an angular distance of less than 40\degr between the flare and Solar Orbiter's footpoint at the SS. This result is broadly consistent with recent statistical studies of impulsive SEP events. For instance, \cite{ho_longitudinal_2024} found that for $^3$He-rich (a widely used indicator of impulsive events) events, the photospheric distance between the flare and the spacecraft footpoint is typically within 30\degr. Similarly, \cite{Warmuth2025} reported that the full widths at half maximum (FWHM) of the longitudinal and latitudinal distributions for impulsive SEEs are 29\degr and 26\degr, respectively. These findings reinforce the longstanding consensus that impulsive SEPs are spatially constrained \citep[e.g.,][]{lin_non-relativistic_1974,reames_particle_1999,ho_upper_2005}. 

Next, we applied the SEA to compare the time profiles of same-side events and opposite-side events.
To avoid contamination from overlapping events, we limited the SEA to isolated events: specifically, we excluded any event for which another event onset occurred within the two hours before or after the onset time. Finally, we have 37 same-side events and 6 opposite-side events for the SEA.

\begin{figure}
\centering
\includegraphics[width=\hsize]{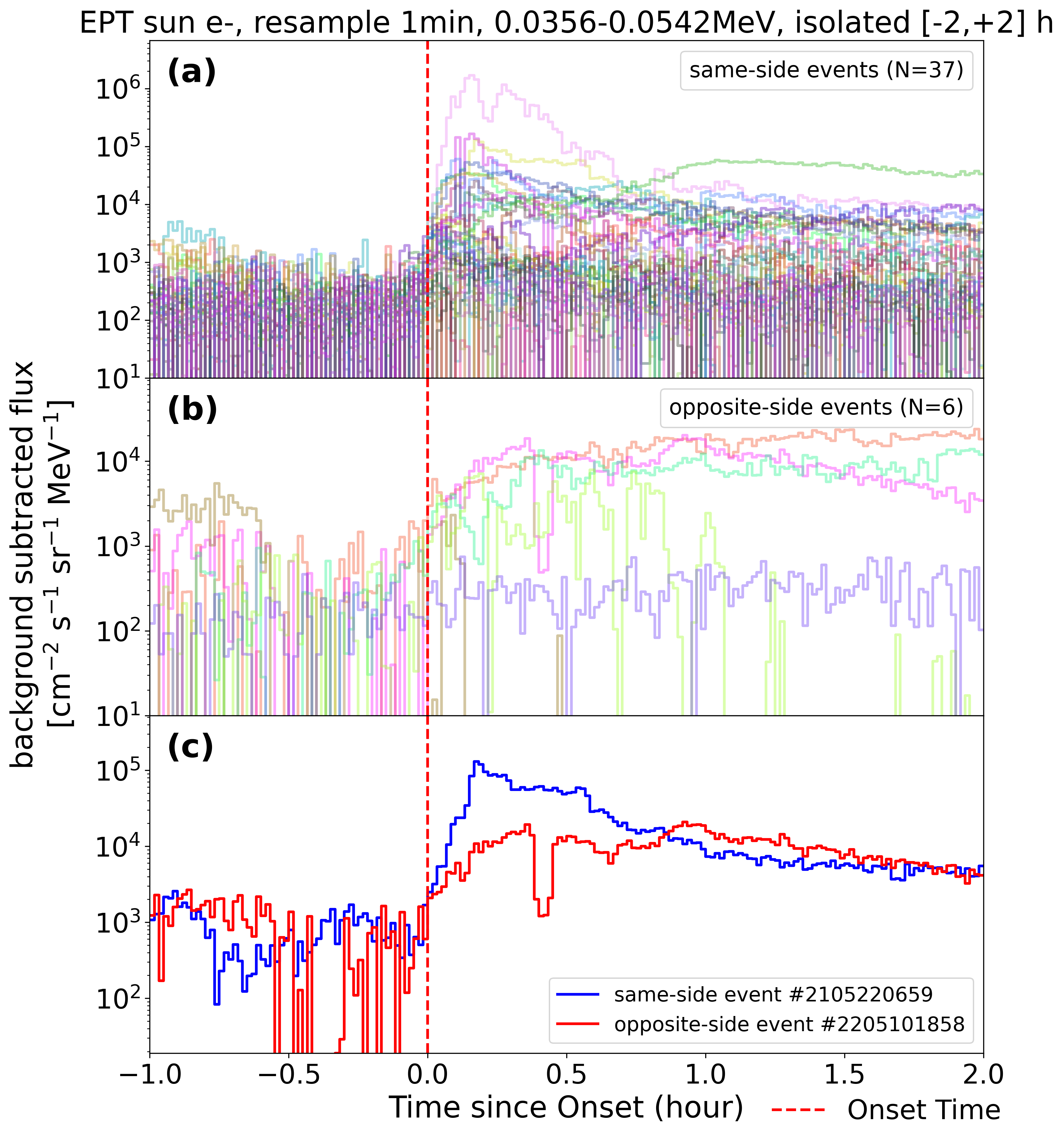}
   \caption{Intensity-time profiles in the energy range of 35.6-54.2 keV detected by the sunward telescope of EPT. Panel (a) shows the profiles of all isolated same-side events, while panel (b) shows those of all isolated opposite-side events. Panel (c) presents two examples of same-side event (blue) and opposite-side event (red). The vertical dashed line marks the event onset time.}\vspace{-5mm}
      \label{time_profile_intensity_HCS_side}
\end{figure}

One SEA focuses on the intensity-time profile. The intensities of events differ by several orders of magnitude, which complicates direct comparison and visualization. Therefore, for each event we subtracted its own background intensity and subsequently added the minimum background intensity in the sample to better display the results on a logarithmic scale. One hour before the onset was chosen to calculate the average background intensity. The SEA results for the intensity-time profiles in 35.6-54.2 keV measured by the sunward sector of EPT are shown in Fig.~\ref{time_profile_intensity_HCS_side}. Figure~\ref{time_profile_intensity_HCS_side} (a) and (b) present the SEA results for the same-side and opposite-side events, respectively. For illustration, Fig.~\ref{time_profile_intensity_HCS_side} (c) shows one representative same-side event and one opposite-side event. The yellow dashed line marks the event onset time.

The intensity-time profiles of SEPs are sensitive to a variety of parameters, such as the solar source, the magnetic connectivity, and the scattering conditions along the transport path \citep[e.g.][]{Wang2022a}. Since the intensity-time profiles are subject to numerous influencing factors, it is challenging to disentangle these effects to solely investigate the impact of the solar source and spacecraft magnetic sector, and attempting to attribute differences in profile morphology solely to the HCS position would be prone to significant uncertainty. Therefore, in this statistical work, we do not undertake further processing or comparison of the intensity-time profiles between the two event categories.

\begin{figure}
\centering
\includegraphics[width=\hsize]{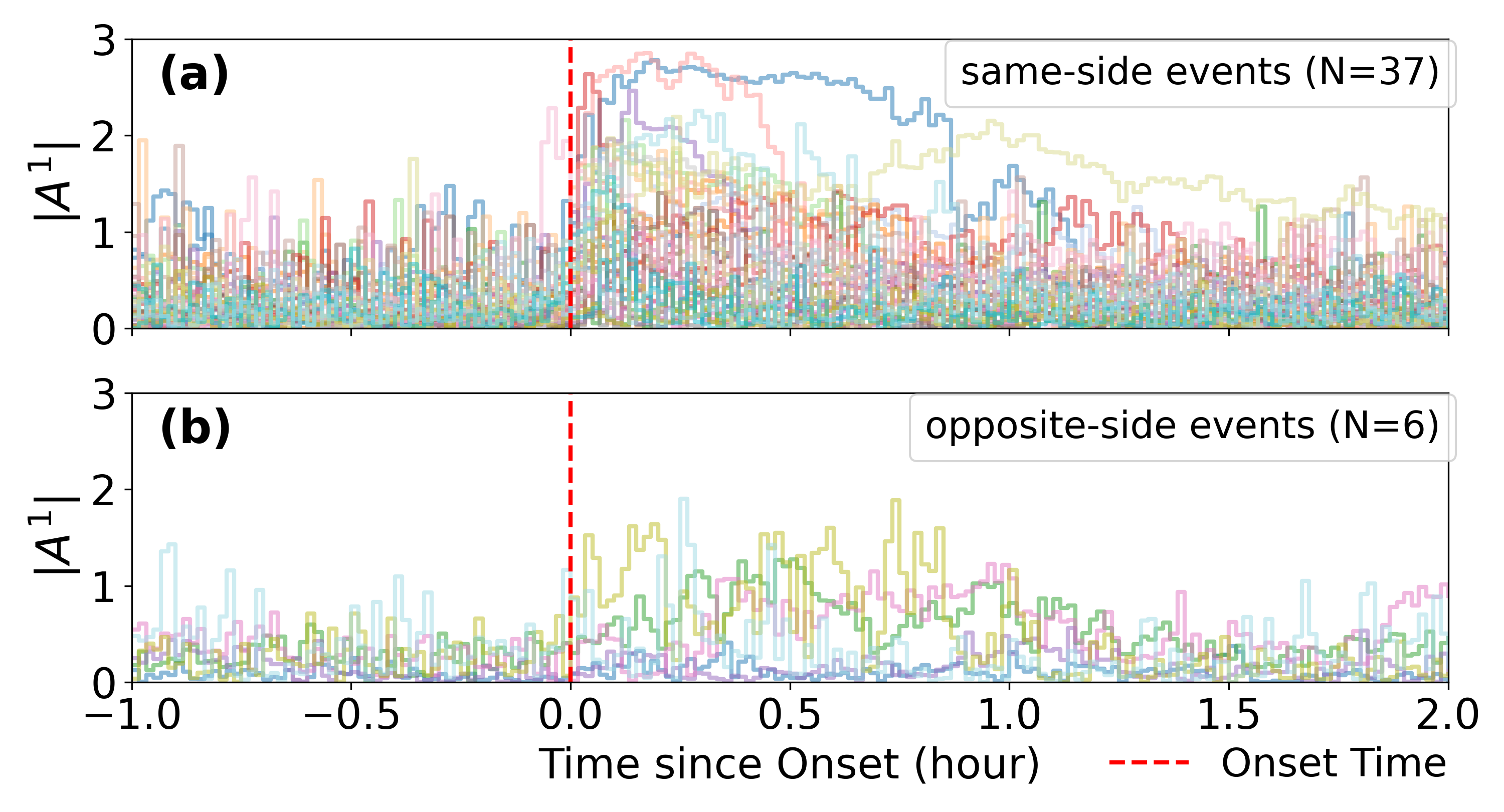}
   \caption{Time profiles of the absolute value of the first-order anisotropy $|A^1|$ of the same-side events (a), and the opposite-side events (b). The first-order anisotropy $A^1$ is calculated using the electron intensities in 35.6-54.2 keV observed by EPT and the local magnetic field vectors by MAG onboard Solar Orbiter, as described in ~\ref{sec:first-order_anisotropy}. The vertical dashed line marks the event onset time.}\vspace{-5mm}
      \label{time_profile_A1_HCS_side}
\end{figure}

The second SEA focuses on the time profile of the absolute value of the first-order anisotropy $|A^1|$. Figure~\ref{time_profile_A1_HCS_side} (a) shows the $|A^1|$-time profile of same-side events, while Fig.~\ref{time_profile_A1_HCS_side} (b) displays those for opposite-side events. The first-order anisotropy $A^1$ is calculated based on the electron intensities in 35.6-54.2 keV observed by EPT. With the limited sample size, the opposite-side events tend to exhibit smaller $|A^1|$ values compared to the same-side events. This result suggests that electrons in opposite-side events may experience stronger scattering during their propagation through the interplanetary medium. However, due to the small number in our sample, this result remains preliminary. Further analysis with a larger dataset is necessary to confirm this trend and to better understand the underlying physical mechanisms influencing particle anisotropy in relation to the HCS position.
\section{Discussion}
\label{Discussion}
\subsection{Physical interpretation of same-side and opposite-side events}
In the SEE sample analysed in this work, the number of same-side events is much greater than that of opposite-side events. If we consider the full flare list, regardless of whether Solar Orbiter detects an SEE event, more opposite-side events are expected. We performed a control analysis using the full Solar Orbiter flare catalogue\footnote{\url{https://github.com/hayesla/stix_flarelist_science}} covering the same time period as the CoSEE-Cat catalogue. The STIX flare list is created by the STIX ground software and data processing team \citep{xiao_data_2023}. For each flare in the list, we selected the three magnetograms closest to the flare peak time as described in \ref{sec:Identification_of_the_solar_source_magnetic_sector} to reconstruct the HCS position via PFSS extrapolation. In the meantime, we determined Solar Orbiter's footpoint at the SS using ballistic back-mapping with the solar-wind speed measured in situ. Based on these locations, we classified the relationship between each flare and Solar Orbiter's footpoint relative to the HCS. Similar to \ref{sec:Identification_of_the_solar_source_magnetic_sector} and \ref{sec:Identification_of_the_spacecraft_magnetic_sector}, we only retained cases where the magnetic polarity determinations are consistent across all three magnetograms. This process yields a total of 3733 cases, comprising 2402 same-side events and 1331 opposite-side events. The result confirms the expectation that there are indeed numerous situations where the flare and Solar Orbiter's footpoint are located on opposite sides of the HCS with large angular distances to the HCS, yet no SEE event is observed.

\begin{figure*}[tbp]
\centering
\includegraphics[width=\textwidth]{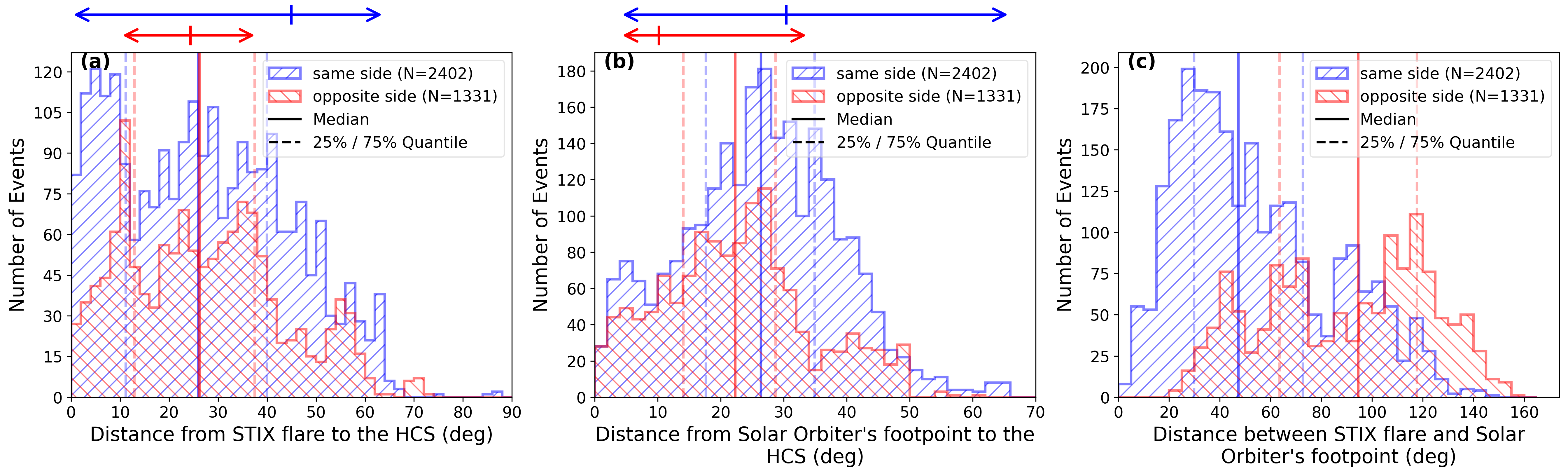}
   \caption{Distributions of distances from flare to the HCS (a), from Solar Orbiter's footpoint to the HCS (b) and from flare to Solar Orbiter's footpoint (c) of the same-side events (blue bars) and opposite-side events (red bars) based on the full STIX flare list. In panels (a) and (b), the horizontal double-headed arrows at the top mark the coverage ranges of the flare-HCS distances and footpoint-HCS distances, respectively, for the same-side (blue) and opposite-side (red) events shown in Fig.~\ref{angle_diff_HCS_side} (a) and (b). The short vertical lines mark the median values of the same-side and opposite-side distributions in Fig.~\ref{angle_diff_HCS_side} (a) and (b).}
      \label{flarelist_angle_diff_HCS_side}
\end{figure*}

Similar to the analysis performed for the SEE events shown in Fig.~\ref{angle_diff_HCS_side}, we calculated the minimum angular distance from the flare and Solar Orbiter's footpoint to the HCS, as well as the separation between flare and footpoint, for the flare list. The results are presented in Fig.~\ref{flarelist_angle_diff_HCS_side}. 
For comparison, the horizontal double-headed arrows at the top of Fig.~\ref{flarelist_angle_diff_HCS_side} (a) and (b) mark the ranges of the flare-HCS and footpoint-HCS distance distributions shown in Fig.~\ref{angle_diff_HCS_side} (a) and (b). The median distances for same-side and opposite-side events in Fig.~\ref{angle_diff_HCS_side} are indicated by short vertical lines. 
Compared with the CoSEE-Cat sample, the median flare--HCS and footpoint--HCS distances of same-side and opposite-side events are more similar in the full flare list. 
The corresponding distributions also show that the same-side and opposite-side groups span comparable ranges. In contrast, in the CoSEE-Cat sample, the opposite-side events are clustered closer to the HCS. 
The absence of opposite-side SEE events with large distances to the HCS may be partly attributable to poor magnetic connectivity to the HCS. 
In other words, particle transport across the HCS is inefficient unless the source or the observer is in close proximity to the HCS. This interpretation aligns with the case study of the 2023 March 13 event by \cite{Waterfall2025}. They found that the spacecraft around the Earth observed the event, with their magnetic connection lying closer to the HCS, whereas BepiColombo and Solar Orbiter did not, with their footpoints located far from the HCS, even though they were closer to the flare than Earth. Our statistical findings reinforce the conclusion of \cite{Waterfall2025} that observer distance from the injection site is not the sole determinant of SEP access. Instead, proximity to the HCS appears to be a dominant factor, providing an effective transport corridor for opposite-side events, particularly when either the source region or the observer is close to the HCS.
\subsection{Robustness tests and sensitivity analyses}
The magnetograms used as boundary conditions for the PFSS model in this study are constructed from observations of the solar disk. Consequently, they inherently represent only the Earth-facing side of the Sun at the time of observation. The far side of the Sun, which is not directly observable from Earth, is typically filled in using flux transport models \citep[e.g.][]{worden_evolving_2000} or synoptic maps \citep[e.g.][]{Sun2018} that compile data over a full solar rotation. This approach introduces a time latency in the magnetogram data, as features on the far side may have evolved significantly by the time they rotate into view.

\begin{figure}
\centering
\includegraphics[width=\hsize]{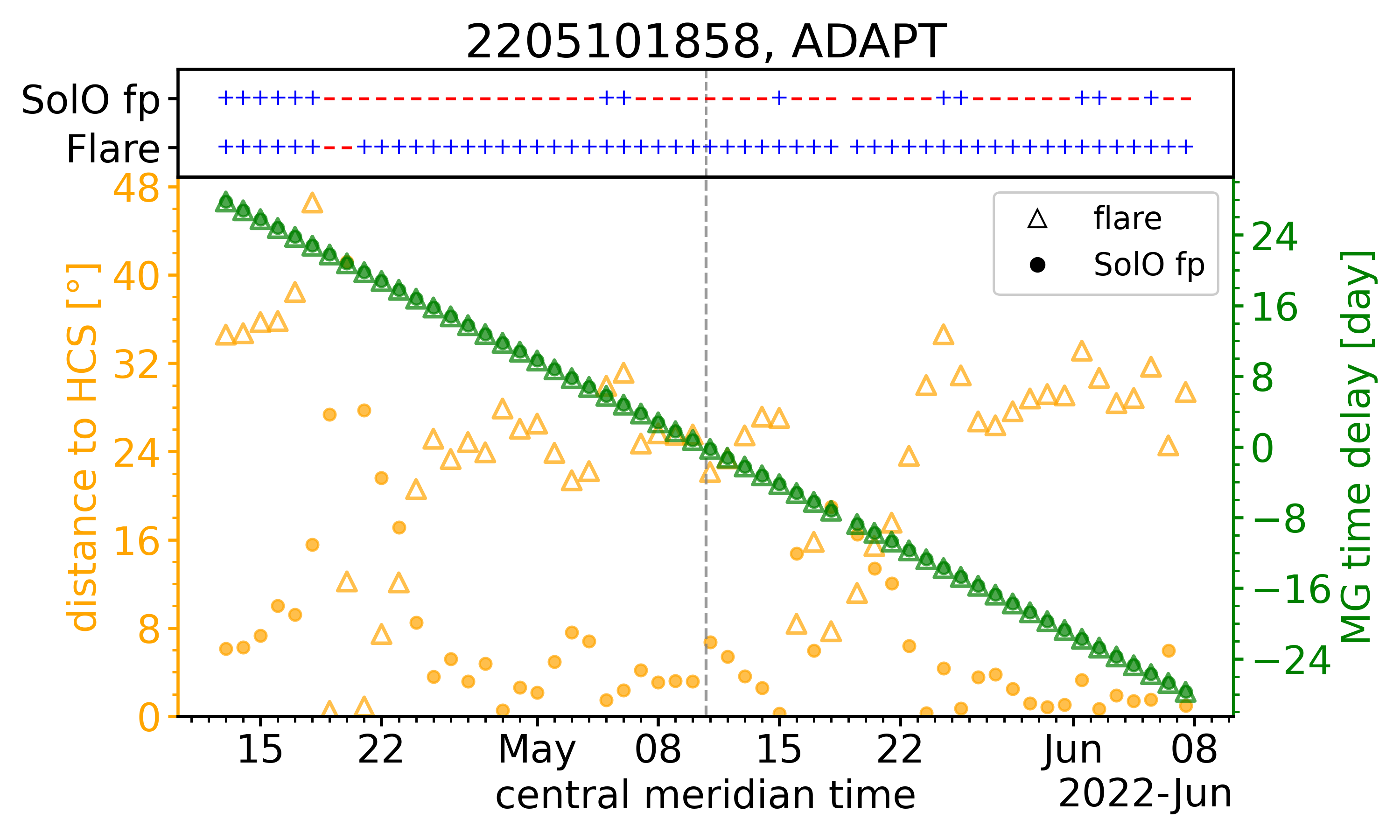}\vspace{-3mm}
   \caption{Results calculated based on multiple magnetograms within $\pm 1$ CR of event \#2205101858. Top panel: Magnetic polarities at the flare location and at Solar Orbiter's footpoint for each magnetogram. Left (orange) axis: Minimum distance to the HCS for the flare (open triangles) and Solar Orbiter's footpoint (filled circles). Right (green) axis: Time latency of the magnetogram data at flare location (Solar Orbiter's footpoint) relative to the flare peak time (SRT). The dashed line marks the SRT of this event.}
      \label{multidata_MGs_example}
\end{figure}

To assess the effect of magnetogram time latency on the determination of whether the flare and the Solar Orbiter's footpoint are on the same or opposite sides of the HCS, we further examined magnetograms within $\pm 1$ CR of each event. For each event and for each magnetogram product (HMI, ADAPT, and GONG), we selected one magnetogram per day within this time window. We then performed PFSS extrapolations, derived the corresponding HCS locations, and determined whether the flare and Solar Orbiter's footpoint are on the same or opposite side of the HCS. Figure~\ref{multidata_MGs_example} shows the result for event \#2205101858. The horizontal axis gives the times corresponding to the CM of the selected magnetograms. The top panel shows the magnetic polarities at the flare location and at Solar Orbiter's footpoint after PFSS extrapolation for each magnetogram. The orange symbols (left-hand axis) show the minimum distance of the flare (open triangles) and Solar Orbiter's footpoint (filled circles) from the HCS. The green symbols (right-hand axis) show the time latency of each magnetogram relative to the flare peak time (open triangles) or the SRT (filled circles). For this event, about 71\% of the tested magnetograms give the same classification as the magnetogram closest in time to the SRT, as described in \ref{sec:Identification_of_the_solar_source_magnetic_sector}.

\begin{figure}
\centering
\includegraphics[width=\hsize]{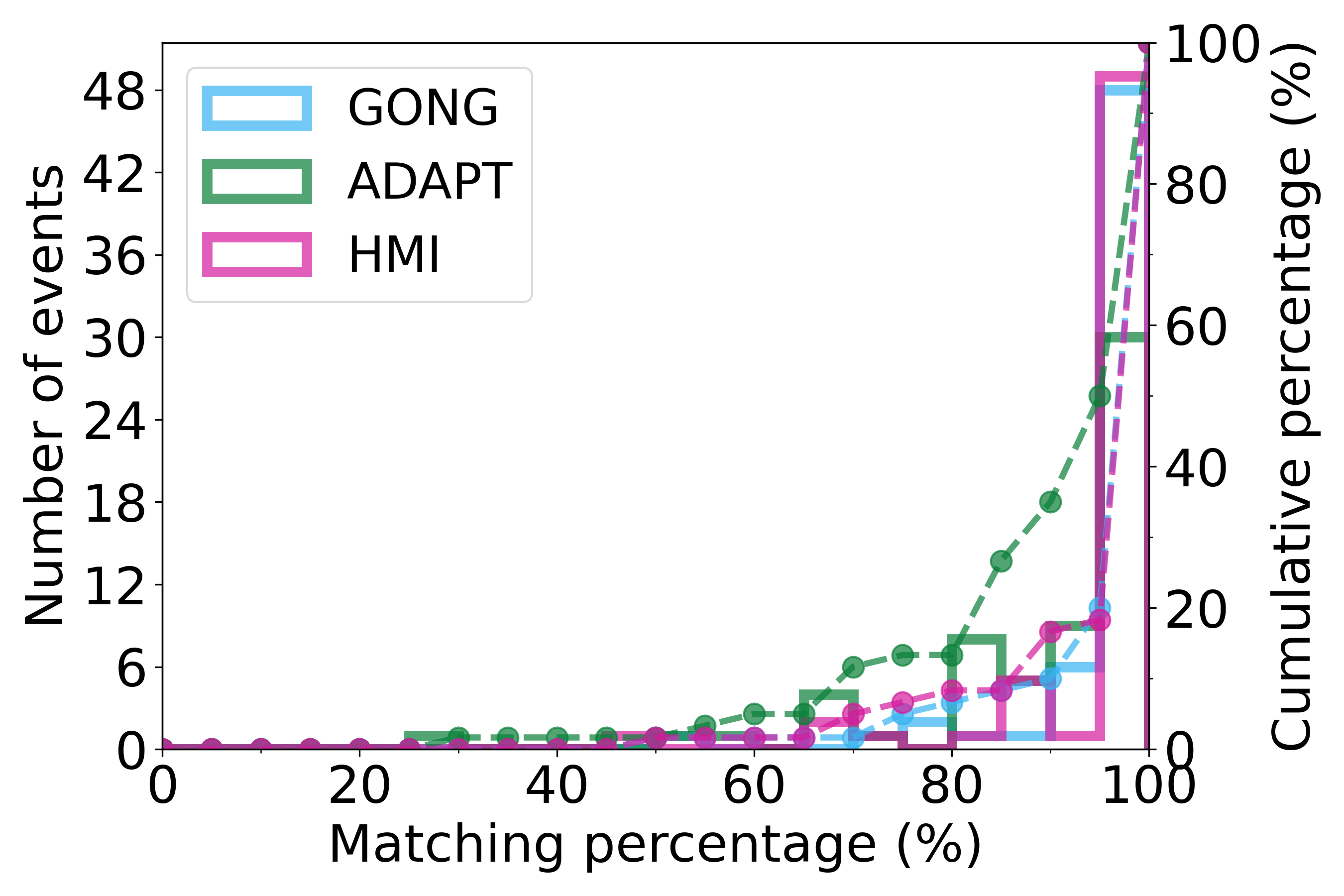}\vspace{-3mm}
   \caption{Distributions of the matching percentages for different magnetogram products (HMI, GONG, and ADAPT), shown in different colours. The histogram (left axis) gives the number of events in each matching percentage bin, and the cumulative curves (right axis, dashed coloured lines) give the fraction of events with matching percentages below a certain value.}\vspace{-4mm}
      \label{matching_percentage_multiMGs}
\end{figure}

We then applied the same analysis to the entire catalogue. For each event, we calculated the percentage of magnetograms within $\pm 1$ CR that give the same classification as the magnetogram nearest to the SRT. We termed this quantity the so-called matching percentage. The overall distributions are shown in Fig.~\ref{matching_percentage_multiMGs}. The left-hand axis shows the number of events in each matching percentage bin. The dashed lines show the cumulative fraction of events with matching percentages below a certain value (right-hand axis). For both HMI and GONG, more than 60\% of the events have matching percentages above 90\%. For ADAPT, roughly half of the events also have matching percentages above 90\%. This result indicates that the same/opposite-side determination is highly consistent even when magnetograms with substantial time latency are used. For those events with same/opposite-side identifications in ~\ref{sec:Event_classification_results}, we list their matching percentages for each magnetogram product in Appendix ~\ref{appendix_event_list}. Overall, within the tested $\pm 1$ CR window, the HCS-based same/opposite-side classification is robust for a large fraction of the sample, although some individual events remain sensitive to magnetogram timing.
\subsection{Outlook and future work}
It is important to note that this study focuses exclusively on electron events in the energy range of tens to hundreds of keV. As demonstrated by the contrasting findings of \cite{Liou2024} and \cite{Waterfall2025}, the influence of the HCS on particle propagation depends critically on many factors, such as specific energy ranges and timescales. Consequently, the HCS may behave differently—acting as a barrier or a conduit—depending on these variables. Further investigations of various scenarios, involving different particle species and energies, are therefore essential to fully clarify the complex and multifaceted role of the HCS in SEP transport.

The time latency of magnetograms can be mitigated in future studies by utilising more advanced magnetograms that incorporate timely far-side observations. Specifically, maps enhanced with helioseismic far-side imaging techniques \citep[e.g.][]{Lindsey2000, Yang2024} and direct magnetic field measurements from the Polarimetric and Helioseismic Imager \citep[PHI;][]{solanki_polarimetric_2020} onboard Solar Orbiter when it is located on the far side of the Sun relative to Earth will significantly improve the global boundary conditions for coronal models \citep[e.g.][]{Yang2023,loeschl_first_2024}. These advancements will contribute to a more accurate reconstruction of the HCS and a better understanding of its role in SEP propagation.

Ultimately, combining improved global magnetic maps with larger multi-spacecraft event samples will be important for testing how strongly the HCS contributes to SEP access and transport.
\section{Conclusions}
In this study, we investigated the influence of the HCS on the transport of SEEs observed by Solar Orbiter. By analysing the sample in the CoSEE-Cat, we classified them into same-side events and opposite-side events based on the relative positions of their associated solar flares with respect to the HCS and Solar Orbiter's magnetic polarity determined by several methods. The spacecraft magnetic polarities determined from source-surface footpoint positions and in situ observations are consistent for most events, with a particularly close agreement among the PFSS+ballistic back-mapping, PAD of strahl electrons, and in situ magnetic field measurements, providing a useful methodological reference for future studies. The uncertainties associated with these classification methods are carefully evaluated. In particular, when same and opposite side events are determined from locations of the HCS, flare, and spacecraft footpoint at the SS, the classification remains robust for a large fraction of events within $\pm 1$ CR window. Following the classification, we identified 60 same-side events and 9 opposite-side events. Our analysis reveals that the opposite-side events tend to have smaller $|A^1|$ values compared to the same-side events. Both the flare and Solar Orbiter's footpoint in opposite-side events tend to be closer to the HCS compared to those in same-side events. This finding suggests that proximity to the HCS is likely a critical factor for magnetic connectivity between the solar source and the spacecraft, as particle transport across the HCS appears inefficient unless either the source or the observer lies close to the HCS. However, given the uncertainties in the PFSS-based reconstruction of the HCS, we cannot completely exclude the possibility that some events classified here as opposite-side events may in fact be same-side events. These results underscore the importance of considering the HCS in understanding SEP propagation and highlight the need for further studies with larger datasets to confirm and expand upon these findings.

\begin{acknowledgements}
Solar Orbiter is a mission of international cooperation between ESA and NASA, operated by ESA. 
This work was supported by the National Natural Science Foundation of China (42225405), the Royal Astronomical Society (RAS), and the Science and Technology Program of Tibet Autonomous Region (XZ202601ZY0001). 
EPD was supported by the German Federal Ministry for Economic Affairs and Energy and the German Space
Agency (Deutsches Zentrum für Luft- und Raumfahrt, e.V., (DLR)), grant number 50OT2002. The UAH team acknowledges the financial support by Project PID2023-150952OB-I00 funded by MICIU/AEI/10.13039/501100011033 and by FEDER, UE. 
HMI data are courtesy of the Joint Science Operations Center (JSOC) Science Data Processing team at Stanford University. 
Data are acquired by GONG instruments operated by NISP/NSO/AURA/NSF with contribution from NOAA. 
This work utilises data produced collaboratively between AFRL/ADAPT and NSO/NISP. 
M. Q. acknowledges the support by the Taishan Scholar Foundation of Shandong Province of China (tsqn202507067). 
C. H. acknowledges the support by the Program of China Scholarship Council. 
The authors would also like to thank Dr. Georgios Nicolaou at University College London and Dr. Yulia Kartavykh at Kiel University for the constructive discussions.
\end{acknowledgements}
\bibliographystyle{bibtex/aa}
\bibliography{bibtex/solar}
\FloatBarrier 
\clearpage
\begin{appendix}
\nolinenumbers
\onecolumn
\section{Event catalogue}
\label{appendix_event_list}
We list the parameters of events with same/opposite-side identification results:
\begin{itemize}
\setlength\itemsep{0.2em}
\setlength\parskip{0pt}
\renewcommand{\labelitemi}{\textendash}
\item \textit{event\_id}: unique identifier of each event in CoSEE-Cat, encodes the EPD onset time in the format YYMMDDHHMM.
\item \textit{side\_mark}: indicates whether Solar Orbiter and the flare are located on the same or opposite side of the HCS; \textit{side\_mark} = 1 denotes the same side, and \textit{side\_mark} = $-1$ denotes opposite sides.
\item \textit{B\_polarity}: magnetic polarity at the Solar Orbiter's footpoint and the flare location; 1 denotes outwards, and $-1$ denotes inwards.
\item \textit{distance\_to\_HCS\_ADAPT}: minimum angular distance between the Solar Orbiter's footpoint / flare location and the HCS derived from PFSS extrapolation using ADAPT magnetograms.
\item \textit{distance\_from\_flare\_STIX\_to\_solofp}: angular distance between the Solar Orbiter's footpoint and the flare location.
\item \textit{MGs\_match\_percentage}: for a given magnetogram (MG) series (GONG/ADAPT/HMI), the percentage of side-classification results (same side vs. opposite side), obtained using magnetograms at different times (within $\pm 1$CR around the event SRT derived from TSA), that are consistent with the reference classification obtained from the magnetogram closest in time to the TSA-derived SRT.
\end{itemize}
\setlength{\LTleft}{0pt}
\setlength{\LTright}{0pt}
\begin{longtable}{@{}llllllllll@{}}
\caption{List of SEE with same/opposite-side identifications.}
\label{tab:cosee_merged_2026_selected_side_mark_pm1}\\
\toprule
\multicolumn{1}{c}{\multirow{2}{*}{\makecell[c]{event\_id\\{}[YYMMDDHHMM]}}} & \multicolumn{1}{c}{\multirow{2}{*}{side\_mark}} & \multicolumn{2}{c}{B\_polarity} & \multicolumn{2}{c}{distance\_to\_HCS\_ADAPT} & \multicolumn{1}{c}{\multirow{2}{*}{\shortstack{distance\_from\\flare\_STIX\\to\_solofp}}} & \multicolumn{3}{c}{MGs\_match\_percentage} \\
\cmidrule(lr){3-4}\cmidrule(lr){5-6}\cmidrule(lr){8-10}
 & \rule{0pt}{2.8ex} & \multicolumn{1}{c}{flare\_STIX} & \multicolumn{1}{c}{solo} & \multicolumn{1}{c}{solofp} & \multicolumn{1}{c}{flare\_STIX} &  & \multicolumn{1}{c}{GONG} & \multicolumn{1}{c}{ADAPT} & \multicolumn{1}{c}{HMI} \\
\midrule
\endfirsthead
\caption[]{List of SEE with same/opposite-side identifications (continued).}\\
\toprule
\multicolumn{1}{c}{\multirow{2}{*}{\makecell[c]{event\_id\\{}[YYMMDDHHMM]}}} & \multicolumn{1}{c}{\multirow{2}{*}{side\_mark}} & \multicolumn{2}{c}{B\_polarity} & \multicolumn{2}{c}{distance\_to\_HCS\_ADAPT} & \multicolumn{1}{c}{\multirow{2}{*}{\shortstack{distance\_from\\flare\_STIX\\to\_solofp}}} & \multicolumn{3}{c}{MGs\_match\_percentage} \\
\cmidrule(lr){3-4}\cmidrule(lr){5-6}\cmidrule(lr){8-10}
 & \rule{0pt}{2.8ex} & \multicolumn{1}{c}{flare\_STIX} & \multicolumn{1}{c}{solo} & \multicolumn{1}{c}{solofp} & \multicolumn{1}{c}{flare\_STIX} &  & \multicolumn{1}{c}{GONG} & \multicolumn{1}{c}{ADAPT} & \multicolumn{1}{c}{HMI} \\
\midrule
\endhead
\midrule
\multicolumn{10}{r}{\small Continued on next page}\\
\endfoot
\bottomrule
\endlastfoot
2104171713 & 1 & -1 & -1 & 5.7 & 13.4 & 70.7 & 76.8 & 80.4 & 68.2 \\
2105220259 & 1 & 1 & 1 & 26.1 & 45.6 & 19.6 & 100 & 83.9 & 100 \\
2105220306 & 1 & 1 & 1 & 26.1 & 45.6 & 19.6 & 100 & 83.9 & 100 \\
2105220345 & 1 & 1 & 1 & 31.5 & 50.9 & 23.0 & 100 & 83.9 & 100 \\
2105220659 & 1 & 1 & 1 & 32.1 & 51.0 & 23.1 & 100 & 83.9 & 100 \\
2105221553 & 1 & 1 & 1 & 25.3 & 45.8 & 25.6 & 100 & 82.1 & 100 \\
2105230456 & 1 & 1 & 1 & 26.3 & 46.4 & 24.0 & 98.2 & 85.7 & 100 \\
2105230940 & 1 & 1 & 1 & 23.2 & 42.9 & 23.6 & 100 & 92.9 & 100 \\
2105231122 & 1 & 1 & 1 & 23.1 & 44.3 & 22.0 & 100 & 94.6 & 100 \\
2105231731 & 1 & 1 & 1 & 23.3 & 45.7 & 22.6 & 100 & 89.3 & 100 \\
2111012155 & -1 & -1 & 1 & 8.5 & 35.3 & 44.4 & 62.5 & 55.4 & 58.7 \\
2112060602 & 1 & 1 & 1 & 23.4 & 1.6 & 44.4 & 73.2 & 85.7 & 74.5 \\
2112311026 & 1 & 1 & 1 & 39.1 & 33.8 & 22.6 & 100 & 100 & 100 \\
2202071958 & 1 & -1 & -1 & 9.4 & 28.1 & 18.7 & 100 & 100 & 100 \\
2202081705 & 1 & -1 & -1 & 14.6 & 29.3 & 17.9 & 100 & 100 & 100 \\
2203052357 & 1 & -1 & -1 & 20.0 & 34.2 & 14.2 & 100 & 100 & 100 \\
2203060804 & 1 & -1 & -1 & 25.5 & 23.0 & 82.1 & 94.6 & 51.8 & 89.6 \\
2203141757 & 1 & -1 & -1 & 32.8 & 26.2 & 63.7 & 96.4 & 98.2 & 95.8 \\
2203302115 & 1 & 1 & 1 & 16.9 & 7.9 & 42.3 & 98.2 & 98.2 & 97.9 \\
2204150302 & 1 & -1 & -1 & 14.1 & 34.6 & 23.8 & 94.6 & 100 & 98.0 \\
2204161900 & 1 & -1 & -1 & 16.3 & 38.1 & 23.6 & 87.5 & 92.9 & 88.0 \\
2204170033 & 1 & -1 & -1 & 12.4 & 40.0 & 27.7 & 80.4 & 80.4 & 80.0 \\
2205080431 & 1 & 1 & 1 & 29.3 & 22.6 & 11.6 & 94.6 & 98.2 & 100 \\
2205080937 & -1 & -1 & 1 & 24.7 & 11.1 & 36.3 & 91.1 & 81.8 & 96.0 \\
2205101858 & -1 & 1 & -1 & 5.7 & 23.1 & 31.2 & 92.9 & 71.4 & 93.9 \\
2205102229 & -1 & 1 & -1 & 4.6 & 22.1 & 28.4 & 91.1 & 57.1 & 87.8 \\
2205110224 & -1 & 1 & -1 & 7.7 & 25.9 & 35.0 & 94.6 & 78.2 & 94.0 \\
2205110353 & -1 & 1 & -1 & 10.2 & 24.4 & 35.9 & 92.9 & 83.6 & 96.0 \\
2205172109 & 1 & -1 & -1 & 12.1 & 22.8 & 30.5 & 91.1 & 92.9 & 93.9 \\
2206182353 & 1 & -1 & -1 & 13.6 & 41.3 & 43.2 & 100 & 94.5 & 97.9 \\
2206202243 & 1 & -1 & -1 & 14.3 & 46.7 & 35.3 & 98.2 & 96.4 & 95.8 \\
2207092110 & 1 & -1 & -1 & 36.8 & 13.0 & 25.8 & 96.4 & 83.9 & 98.1 \\
2207231943 & 1 & 1 & 1 & 4.4 & 37.1 & 32.8 & 100 & 72.7 & 100 \\
2207300857 & -1 & -1 & 1 & 16.4 & 36.8 & 85.4 & 100 & 100 & 100 \\
2209221327 & 1 & 1 & 1 & 42.8 & 46.1 & 34.9 & 98.2 & 100 & 97.9 \\
2209271028 & 1 & -1 & -1 & 19.8 & 6.8 & 37.1 & 98.2 & 58.2 & 89.4 \\
2209280525 & 1 & -1 & -1 & 23.1 & 10.3 & 40.6 & 98.2 & 65.5 & 95.7 \\
2209291228 & 1 & -1 & -1 & 35.6 & 6.9 & 53.8 & 98.2 & 65.5 & 95.7 \\
2210270948 & 1 & 1 & 1 & 28.8 & 52.7 & 27.2 & 100 & 100 & 100 \\
2210281744 & 1 & 1 & 1 & 33.0 & 24.7 & 61.0 & 50.0 & 89.3 & 66.7 \\
2211110026 & 1 & -1 & -1 & 58.0 & 62.9 & 5.4 & 100 & 98.2 & 100 \\
2211110143 & 1 & -1 & -1 & 58.2 & 63.5 & 5.5 & 100 & 96.4 & 100 \\
2211110155 & 1 & -1 & -1 & 58.2 & 63.2 & 5.2 & 100 & 96.4 & 100 \\
2211110324 & 1 & -1 & -1 & 64.9 & 60.3 & 4.8 & 100 & 92.9 & 100 \\
2211110718 & 1 & -1 & -1 & 63.6 & 60.8 & 5.0 & 100 & 98.2 & 100 \\
2211110949 & 1 & -1 & -1 & 41.5 & 45.7 & 4.2 & 100 & 96.4 & 100 \\
2211111141 & 1 & -1 & -1 & 41.9 & 46.5 & 5.3 & 100 & 96.4 & 100 \\
2211111400 & 1 & -1 & -1 & 41.9 & 45.8 & 4.4 & 100 & 96.4 & 100 \\
2211111819 & 1 & -1 & -1 & 42.3 & 45.9 & 3.7 & 100 & 92.9 & 100 \\
2211120231 & 1 & -1 & -1 & 61.5 & 63.8 & 3.5 & 100 & 96.4 & 100 \\
2211120442 & 1 & -1 & -1 & 64.6 & 61.1 & 3.6 & 100 & 98.2 & 100 \\
2211121030 & 1 & -1 & -1 & 55.9 & 49.2 & 6.7 & 100 & 96.4 & 100 \\
2211121257 & 1 & -1 & -1 & 55.9 & 47.5 & 8.8 & 100 & 96.4 & 100 \\
2211121745 & 1 & -1 & -1 & 55.2 & 40.3 & 19.4 & 98.2 & 91.1 & 95.7 \\
2211121757 & 1 & -1 & -1 & 55.3 & 52.2 & 4.6 & 100 & 94.6 & 100 \\
2211121806 & 1 & -1 & -1 & 55.3 & 50.5 & 5.7 & 100 & 96.4 & 100 \\
2211131314 & 1 & -1 & -1 & 11.2 & 23.0 & 19.1 & 91.1 & 89.3 & 89.4 \\
2211191309 & 1 & -1 & -1 & 15.3 & 5.1 & 19.0 & 92.9 & 67.9 & 89.4 \\
2212010558 & 1 & 1 & 1 & 26.9 & 23.1 & 99.3 & 75.0 & 29.6 & 45.8 \\
2212010724 & 1 & 1 & 1 & 27.2 & 40.2 & 30.1 & 100 & 100 & 100 \\
2212111148 & 1 & -1 & -1 & 65.0 & 41.3 & 36.8 & 100 & 65.4 & 95.8 \\
2212210225 & 1 & 1 & 1 & 16.2 & 19.2 & 117.3 & 100 & 100 & 100 \\
2212230458 & 1 & 1 & 1 & 35.1 & 53.5 & 21.6 & 100 & 98.0 & 100 \\
2212240116 & 1 & 1 & 1 & 36.4 & 57.9 & 38.1 & 100 & 98.0 & 100 \\
2212240416 & 1 & 1 & 1 & 34.8 & 55.8 & 26.6 & 100 & 98.0 & 100 \\
2212242245 & -1 & -1 & 1 & 32.1 & 33.7 & 158.1 & 100 & 98.0 & 100 \\
2212250155 & 1 & 1 & 1 & 40.9 & 58.6 & 28.7 & 100 & 98.0 & 100 \\
2212280450 & 1 & 1 & 1 & 49.7 & 60.6 & 20.1 & 100 & 98.0 & 100 \\
2212301617 & -1 & -1 & 1 & 20.6 & 11.3 & 34.2 & 100 & 83.7 & 98.0 \\
\end{longtable}
\begin{multicols}{2}
\raggedcolumns
\section{Methodological uncertainties and limitations}
\label{appendix_Methodological_uncertainties_and_limitations}
The uncertainties discussed below arise at different stages of the analysis, from solar source identification to magnetic mapping and polarity inference.

Solar flares associated with the electron events in the CoSEE-Cat catalogue are primarily investigated using STIX, with supplementary contextual information on associated flares and eruptive phenomena provided by Extreme-Ultraviolet Imager (EUI). The identification of solar sources for these electron events is subject to several uncertainties. STIX and EUI observations sometimes reveal multiple flares occurring in close temporal or spatial proximity, making it difficult to unambiguously associate a single flare with a given event. The STIX flare selected as the solar source in the CoSEE-Cat catalogue is typically the one closest to the SEE solar release time, with its peak occurring before the EPD onset. However, the solar release time is inferred from the EPD onset, which itself can be uncertain, especially in cases where multiple events overlap and the onset of each individual event is difficult to determine. In certain other cases, the true flare may be located behind the limb as seen from Solar Orbiter and thus not observable, leading to apparent flare-less events. For widespread SEP events, the identification of the solar source region becomes even more challenging due to the broad spatial extent and possible contributions from multiple active regions. In such cases, the original flare association in the catalogue is revised manually to account for ambiguous or uncertain solar origins. As already noted in the CoSEE-Cat catalogue \citep{Warmuth2025}, some SEE events in the list may have incorrectly assigned parent solar sources and would require further detailed analysis.

Beyond source association itself, additional uncertainty arises from the positional accuracy of the selected flare. 
The flare locations at the photosphere provided in the CoSEE-Cat catalogue are reconstructed using the EM imaging algorithm adapted for STIX \citep{massa_count-based_2019}. This indirect imaging method may yield unreliable results, particularly when two or more active regions flare simultaneously in different areas of the solar disc—a situation that becomes more frequent with increasing solar activity. Additionally, projection effects for flares occurring near the solar limb can distort their apparent position when mapped to the photosphere. These factors collectively contribute to uncertainties in determining the flare location at the photosphere.

Even when the photospheric flare position is reasonably constrained, mapping that location to the SS remains challenging. 
We note that the flare location at the photosphere differs from its footpoint position at the SS due to the complex magnetic field structures in the corona. To address this, we attempted to perform PFSS for each flare to obtain its footpoints at the SS with the three magnetograms described in \ref{sec:Identification_of_the_solar_source_magnetic_sector}. 
In practice, however, PFSS-based flare mapping is not straightforward. When tracing magnetic field lines directly from the flare location at the photosphere, some flares do not connect to open field lines at all. 
We therefore expanded the mapping region by placing 500 seed points on the photosphere within a spherical angular radius of 5\degr\ around each flare. The 5\degr radius corresponds to a circular area on the photosphere with a central angle of 10\degr, which is comparable to the typical size of flare active regions, as reported in the statistical study \citep{Kazachenko2017}. 
An example of the PFSS mapping result for event \#2105220306 is shown in Fig.~\ref{flare_PFSS_mapping_example}. The ADAPT magnetogram is used as the boundary condition for PFSS in this example. The flare location at the photosphere is marked by a star. The seeds around the flare at the photosphere are shown as green dots. The corresponding footpoints at the SS are shown as grey dots, and the HCS is indicated by the blue line. 
Despite this expanded seeding approach, only a fraction of the 205 STIX flares are found to connect to open field lines: 80/205 for GONG, 94/205 for ADAPT, and 42/205 for HMI. For the subset of flares with open field lines, all footpoints at the SS generally fall in the same magnetic sector as the flare position obtained by radial projection from the photosphere, with full agreement (100\%) for GONG and ADAPT and a lower agreement of 64\% for HMI. 
These results should also be viewed in the context of the known limitations of PFSS-based mapping. 
The PFSS assumes a current-free corona between the photosphere and a spherical SS, and therefore cannot represent electric currents, non-potential fields, or time-dependent transients. Its solutions furthermore depend strongly on the height and shape of the SS \citep{2020A&A...638A.109K,2021A&A...645A..83K}, as well as on other model parameters \citep{Luhmann2012,Koukras2025}. 
As a result, the PFSS-derived HCS traces and footpoints should be treated as approximate, model-dependent estimates. In this study, we partly mitigated these limitations by comparing HMI, GONG, and ADAPT magnetograms. Although these magnetograms differ in native spatial resolution, centre-to-limb sampling, and polar-field treatment, previous studies have shown broadly similar large-scale HCS structures and only moderate model-dependent shifts in the inferred HCS position \citep[e.g.][]{Li2021,Perri2023}. More generally, model uncertainties are expected to broaden the range of possible source locations rather than systematically displacing them to very different locations, and their influence should therefore be stronger for events close to the HCS than for those already well separated from it \citep[e.g.][]{Koukras2025}. For analyses requiring higher fidelity, nonlinear force-free or time-dependent Magnetohydrodynamic (MHD)/data-driven models are preferable despite their greater data and computational cost, and may be particularly helpful for better distinguishing events located close to the HCS. Taking these considerations together, we used the radially
\begin{center}
\begin{minipage}{\columnwidth}
\centering
\includegraphics[width=\hsize]{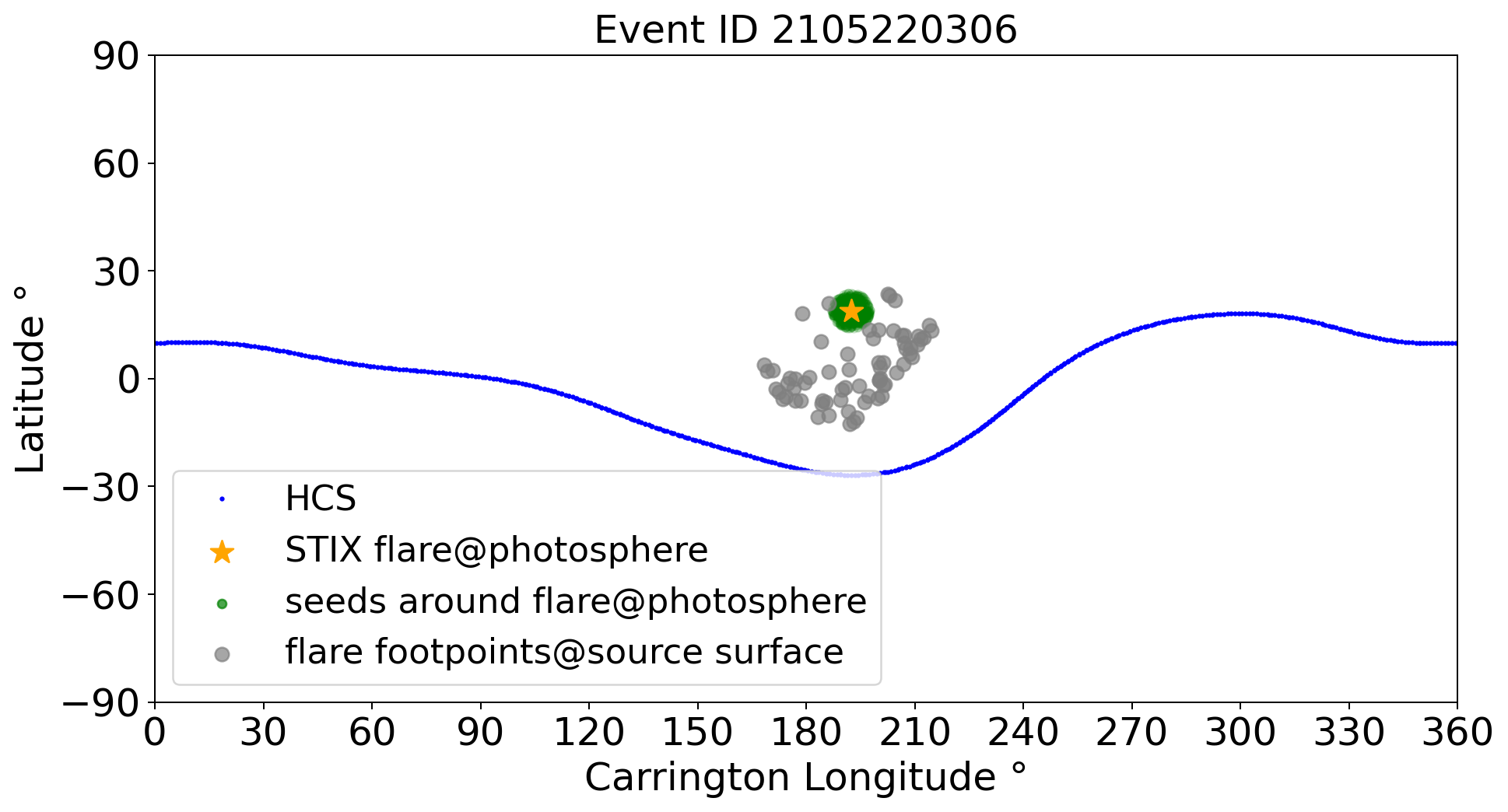}
\captionof{figure}{PFSS magnetic mapping result of the STIX flare in event \#2105220306. The HCS is indicated by the blue line. The STIX flare at the photosphere is indicated by the red star. The mapping seeds around the flare at the photosphere at the start of mapping are indicated by the green dots. Footpoints at the SS of the seeds are marked by grey dots.}\vspace{0mm}
\label{flare_PFSS_mapping_example}
\end{minipage}
\end{center}
\columnbreak
projected flare location to determine the flare magnetic sector in this work. This approximation is also quite common in related studies, in which the flare--spacecraft footpoint separation angle is estimated from the flare location at the photosphere and the back-mapping spacecraft footpoint at the SS \citep[e.g.][]{Lario2006,zhao_influence_2007,Park2024,Dresing2024}.

An additional source of uncertainty arises from the ballistic back-mapping, which assumes a nominal Parker spiral and a predominantly radial solar wind. The large-scale heliospheric magnetic connectivity can be more complex than implied by a simple Parker-spiral backmapping \citep[e.g.,][]{fisk1996motion}. At the same time, some simulations suggest that latitudinal transport is less efficient than longitudinal transport in a Parker-spiral geometry with large-scale magnetic fluctuations \cite[e.g.][]{kelly2012cross}. Thus, the resulting latitudinal error is not expected to systematically dominate the angular-distance estimates.

On the spacecraft side, local measurements may not always provide a unique sector classification. 
Although the strahl electron PAD provides a convenient proxy for the local magnetic polarity and reduces sensitivity to short timescale magnetic fluctuations, several caveats remain. Magnetic structures that connect regions of opposite polarity over a range of scales, such as flux ropes, can produce simultaneous parallel and antiparallel (counterstreaming) strahl populations \citep[e.g.][]{Fedorov2021}. Such bidirectional strahl associated with complex topologies therefore produces ambiguous polarity signatures and complicates the interpretation of the local magnetic polarity. Additionally, local particle acceleration and wave-particle interactions (e.g. whistler-mode scattering) can broaden or redirect the strahl, and shocks, stream interaction regions or other transient compressions can substantially modify the PAD. Taken together, these factors contribute to the uncertainty in determining the spacecraft magnetic sector.

The large-scale morphology of the HCS is also not fixed in time. Its waviness and inclination relative to the ecliptic vary with solar-cycle phase, and these changes can influence drift-related particle transport. In the present study, however, the SEE sample is restricted to the CoSEE-Cat interval from November 2020 to December 2022, that is, a relatively limited time span during the rise toward solar maximum. The sample therefore does not cover the full range of HCS configurations expected over an entire solar cycle. Even within this interval, event-to-event differences in the large-scale HCS geometry may still contribute to the spread in the observed properties. We also acknowledge that the inclination of the HCS and effective thickness along the electrons path may affect sector assignment, particularly for events close to the HCS. However, such potentially ambiguous cases were largely excluded by our selection procedure, making the final sample relatively robust in this respect.

Taken together, these sources of uncertainty can affect individual events, but the multi-method consistency criteria in this study are intended to reduce their impact on the overall statistical conclusions.
\end{multicols}
\end{appendix}
\end{document}